\documentclass[a4paper,fleqn]{cas-dc}

\usepackage[numbers]{natbib}

\usepackage{amsmath,amsthm}
\usepackage{dsfont}
\usepackage{amssymb}

\usepackage[utf8]{inputenc}
\usepackage[T1]{fontenc}
\usepackage{booktabs}

\usepackage{subcaption}

\begin{document}
\let\WriteBookmarks\relax
\def\floatpagepagefraction{1}
\def\textpagefraction{.001}
\shorttitle{The strong effect of network resolution on energy system models}
\shortauthors{Martha Maria et~al.}
%\begin{frontmatter}

\title [mode = title]{The strong effect of network resolution on electricity system models with high shares of wind and solar}
\tnotemark[1]

\tnotetext[1]{This document is the results of the research
   project funded by Helmholtz Association under grant no. VH-NG-1352.}

\author[kit,cor1]{Martha Maria Frysztacki}
\fntext[cor1]{Corresponding author \texttt{martha.frysztacki@kit.edu} (Martha Maria)}
\credit{Conceptualization, Methodology, Software, Formal Analysis, Data Curation, Writing - Original Draft, Writing - Review \& Editing, Visualization}

\author[kit,fias]{Jonas Hörsch}
\credit{Conceptualization, Methodology, Software, Data Curation, Writing - Review \& Editing, Visualization}

\author[kit]{Veit Hagenmeyer}
\credit{Writing - Review \& Editing, Funding Acquisition}

\author[kit,fias]{Tom Brown}
\credit{Conceptualization, Methodology, Writing - Original Draft, Writing - Review \& Editing, Project Administration, Funding Acquistion}

\address[kit]{Institute for Automation and Applied Informatics, Karlsruhe Institute of Technology, 76344 Eggenstein-Leopoldshafen, Germany}

\address[fias]{Frankfurt Institute for Advanced Studies, Ruth-Moufang-Straße 1, 60438 Frankfurt am Main, Germany}

\begin{abstract}
Energy system modellers typically choose a low spatial resolution
for their models based on administrative boundaries such as
countries, which eases data collection and reduces computation
times. However, a low spatial resolution can lead to sub-optimal
investment decisions for wind and solar generation. Ignoring power grid
bottlenecks within regions tends to underestimate system costs,
while combining locations with different wind and solar capacity
factors in the same resource class tends to overestimate costs.
We
investigate these two competing effects in a capacity expansion model for  Europe's power system with a high share of
renewables,  taking advantage of newly-available high-resolution datasets as well as computational advances. We vary the number of nodes, interpolating between a
37-node model based on country and synchronous zone boundaries, and
a 1024-node model based on the location of electricity
substations. If we focus on the effect of renewable resource
resolution and ignore network restrictions, we find that a higher
resolution allows the optimal solution to concentrate wind and solar capacity
at sites with better capacity factors and thus reduces system costs by up to 10\% compared to a  %1 - 222.459564/248.112065 = 0.10339078432159277
low resolution model. This results in a big swing from offshore to
onshore wind investment. However, if we introduce grid bottlenecks
by raising the network resolution, costs increase by up to 23\% as  % 272.876699/222.459564 = 1.2266350526516359
generation has to be sourced more locally at sites with worse
capacity factors.
%As a result, the optimal solution with a high number of nodes
%for both renewable resources and the network can be up to 9\% more   %270/248 = 1.09
%expensive than with a low number of nodes, as well as leading to
%different investment choices.
These effects are most pronounced in scenarios where
grid expansion is limited, for example, by low local acceptance. We show that allowing grid expansion
mitigates some of the effects of the low grid resolution, and lowers
overall costs by around 16\%.  % 1 - 228.761924/272.876699 = 0.161665588
\end{abstract}

%\includegraphics{figs/grabs.pdf}

\begin{highlights}
\item Highly-renewable European power system is optimized at high spatial resolution
\item High-resolution capacity placement for wind and solar reduces costs by up to 10\%
\item Models with low network resolution ignore congestion, underestimating costs by 23\%
\item Costs underestimated most when grid expansion limited by, e.g., public acceptance
\item Grid reinforcements relieve congestion and lower system costs by up to 16\%
\end{highlights}


\maketitle

\section{Introduction}

Electricity systems with high shares of wind and solar photovoltaic
generation require a fundamentally different kind of modelling to
conventional power systems with only dispatchable
generation \cite{Pfenninger2014}. While investments in conventional
power plants can be dimensioned according to simple heuristics like
screening curves \cite{biggar2014}, the assessment of wind and solar resources requires
a high temporal and spatial resolution to capture their weather-driven
variability.  The need to assess investments in generation,
transmission and flexibility options over thousands of representative
weather and demand situations, as well as over thousands of potential
locations, means that balancing model accuracy against computational
resources has become a critical challenge.

The effects of temporal resolution have been well researched in the
electricity system planning literature \cite{burdenresponse},
including the need for at least hourly modelling resolution
\cite{Pfenninger2014}, the consequences of clustering representative
conditions \cite{KOTZUR2018474}, and the need to include extreme
weather events \cite{Perera2020}. On the spatial side, it has been
recognized that integrating renewable resources on a continental scale
can smooth large-scale weather variations, particularly from wind
\cite{Czisch2005}, and avoid the need for temporal balancing.  This
smoothing effect has been found in studies of the benefits of grid
expansion both in Europe, where the impact on balancing needs
\cite{Rodriguez2013} and storage requirements
\cite{Schlachtberger2017} has been analysed, and in the United States \cite{macdonald2017}.
However, there has been little research on the effects of spatial
resolutions on planning results. This is partly due to the fact that
collecting high-resolution spatial data is challenging, as well as the
fact that optimization at high-resolution over large areas is
computationally demanding.

Choosing the spatial resolution based on administrative boundaries
such as country borders --which is a common approach in the literature \cite{Czisch2005,Rodriguez2013,GILS2017173}--
fails to account for the variation of resources inside large countries
like Germany. Aggregating low-yield sites together with high-yield
sites takes away the opportunity to optimize generation
placement, which distorts investment decisions and drives up costs.

On the other hand, aggregating diverse resources to single points
tends to underestimate network-related costs, since the models are
blind to network bottlenecks that might hinder the welfare-enhancing
integration of renewable resources located far from demand centers.
The effects of network restrictions are all the more important given
the apparent low public acceptance for new overhead transmission
lines, observed in Germany \cite{GALVIN2018114} and across Europe \cite{COHEN20144}, and the long planning and
construction times for new grid infrastructure \cite{TYNDP2018}.

In the present contribution we introduce a novel methodology to
disentangle these two competing spatial effects of resource and
network resolution, so that for the first time their different impacts
on system costs and technology choices can be quantified. We then demonstrate the methodology by running simulations in a model of the future European electricity system with a higher spatial resolution than has previously been achieved in the literature. We optimize investments and operation of
generation, storage and transmission jointly in a system with a high share of renewables under a 95\% reduction in CO$_2$ emissions compared to 1990,
which is consistent with European targets for 2050 \cite{2050roadmap}.
A recently-developed, high-resolution, open-source model of the European transmission network,
PyPSA-Eur \cite{PyPSA-Eur}, is sequentially clustered from 1024 nodes
down to 37 nodes in order to examine the effects on optimal investments in
generation, transmission and storage.

Previous work in the engineering literature has focused on the effect
of different network clustering algorithms \cite{Jain1999} on the
flows in single power flow simulations \cite{Blumsack2009,Hamon2015},
or used clustering algorithms that are dependent on specific dispatch
situations \cite{Cheng2005,Singh2005,Shayestah2015} and therefore
unsuitable when making large changes to generation and transmission
capacities. In the planning literature that considers a high share of renewables in the future energy system, the effects of clustering applied separately to wind, solar and demand were
investigated in \cite{Siala2019}, but neglected potential transmission line congestion within large regions. In \cite{Kueppers2020} the previous study was extended by including a synthesized grid and renewable profiles, but it ignored the existing topology of the transmission grid. Effects of varying the
resolution were not considered in either of the studies. Recent work has examined regional solutions for the European power system, but did not take
into account existing transmission lines, potential low public
acceptance for grid reinforcement or the grid flow physics \cite{Troendle2020}. Other studies have examined transmission grid expansion at substation resolution, but either the temporal resolution was too low to account for wind and solar
variability \cite{Egerer2013,Hess2018}, or only single countries were
considered \cite{Nabe2014,KWK2,Hess2018}, or transmission expansion
was not co-optimized with generation and storage \cite{Egerer2013,Brown2016,Schaefer2017}. The competing effect of clustering transmission lines versus variable resource sites on the share of renewables was also discussed in \cite{Cole2017}, but the report did not provide an analysis of how strongly the respective clustering impacts modeling and planning results. The effects of model resolution on system planning results were considered for the United States in \cite{krishnan2016}, where a cost-benefit was seen for higher wind and solar resolution, but the resource resolution was not separated from the network resolution, and only a small number of time slices were considered to represent weather variations.

Advances in solver algorithms and code optimization in the modelling
framework PyPSA \cite{PyPSA}, as well as hardware improvements, allow
us to achieve what was previously not possible in the literature: the
co-optimization of transmission, generation and storage at high
temporal and spatial resolution across the whole of Europe, while taking into account
linearized grid physics, existing transmission lines and realistic
restrictions on grid reinforcement.  In previous work by some of the
authors large effects of spatial resolution on investment results were
seen \cite{Hoersch2017}, but because the resource and network
resolution were changed in tandem, it was not possible to analyse
which effect dominates the results. In the present contribution we
present a novel study design that separates the effects of resource and network resolution, and
demonstrate the substantial differences between the two effects using the high-resolution simulations enabled by recent software and hardware advances.

\section{Methods} \label{sec:matherialandmethods}

In this section we present an overview of the underlying model and the
study design, before providing more details on the clustering
methodology and the investment optimisation. A list of notation is provided in Table
\ref{tab:notation}.

\subsection{Model input data}

\begin{figure}[!t] \centering
	\includegraphics[width=9cm]{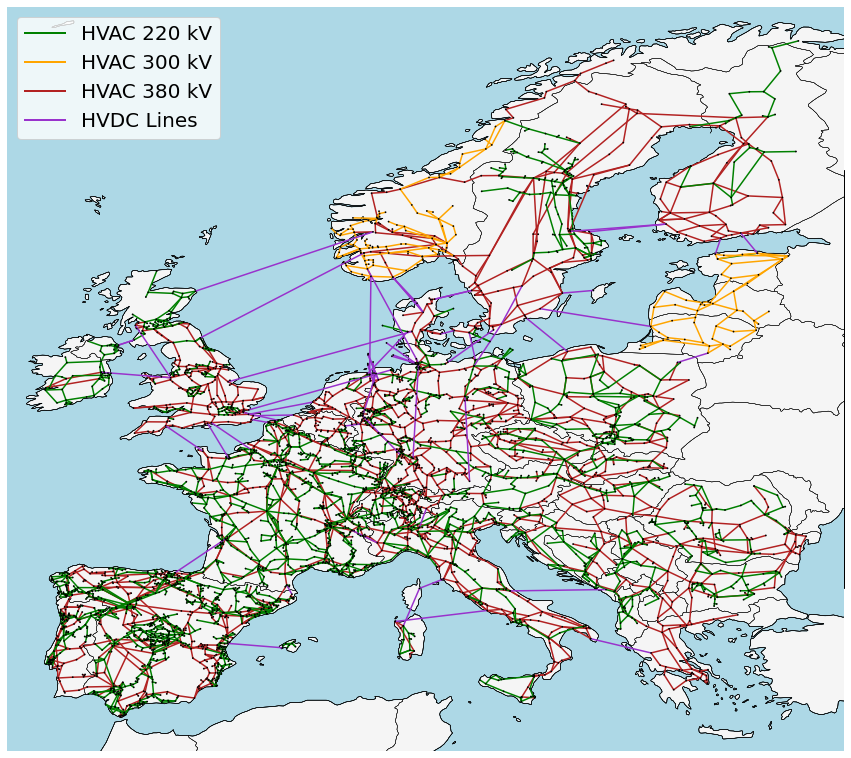}
	\caption{PyPSA-Eur model of the European electricity system, including all existing and planned high-voltage alternating current (HVAC) and direct current (HVDC) lines.}
	\label{fig:europe-map}
\end{figure}

The study is performed in a model of the European electricity system
at the transmission level, PyPSA-Eur, which is fully described in a
separate publication \cite{PyPSA-Eur}. Here we give a brief outline of
the input data.

The PyPSA-Eur model shown in Figure \ref{fig:europe-map} contains all
existing high-voltage alternating current (HVAC) and direct current
(HVDC) lines in the European system, as well as those planned by the
European Network of Transmission System Operators for Electricity
(ENTSO-E) in the Ten Year Network Development Plan
(TYNDP) \cite{TYNDP2018}. The network
topology and electrical parameters are derived from the ENTSO-E
interactive map \cite{interactive} using a power grid extraction toolkit \cite{wiegmans_2016_55853}. In total the network consists of 4973 nodes, 5721 HVAC and 32 HVDC lines existing as of 2018, as well as 279 HVAC and 29 HVDC planned lines.

Historical
hourly load data for each country are taken from the Open Power System
Data project \cite{OPSD} and distributed to the nodes within each
country according to population and gross domestic product data.
Generation time series are provided for the
surrounding wind and solar plants based on historical wind and
insolation data derived from the ERA5
reanalysis dataset \cite{ERA5} and the SARAH2 surface radiation
dataset \cite{SARAH2}. Renewable installation potentials are based on land
cover maps, excluding for example nature reserves, cities or streets.

The model was partially validated in \cite{PyPSA-Eur}. Further
validation against historical data was carried out in
\cite{Frysztacki2020}, where it was shown that the model could
reproduce curtailment of wind and solar in Germany due to transmission
bottlenecks in the years 2013-2018. The ability to reproduce historical
congestion provides a strong check on the match between the transmission
network data and the availability of wind and solar generation in the model.

\subsection{Clustering study design}

The nodes of the model are successively clustered in space into a
smaller number of representative nodes using the $k$-means
algorithm \cite{Hartigan1979}.  This groups close-by nodes together, so
that, for example, multiple nodes representing a single city are
merged into one node. Nodes from different countries or different
synchronous zones are not allowed to be merged; to achieve this, the
overall number of desired nodes is partitioned between the countries
and synchronous zones before the $k$-means algorithm is applied in each
partition separately. In total there are 37 `country-zones' in the
model, i.e. regions of countries belonging to separate synchronous
zones.

Figure \ref{fig:clustered-methodology}, Case 1
shows the results for Ireland and the United Kingdom (where Northern
Ireland is in a separate synchronous zone to Great Britain). Once the
nodes have been clustered, they are reconnected with transmission
corridors representing the major transmission lines from the
high-resolution model. Electricity demand, conventional generation and storage options are also aggregated to the nearest network node. More technical details on the clustering
can be found in subsection \ref{sec:clusteringdetails}. An analysis of the effects of clustering
on the network flows can be found in the Appendix, Section \ref{sec:flows}.

\subsection{Resource versus network resolution case studies}

\begin{figure*} \centering
	\includegraphics[width=\textwidth]{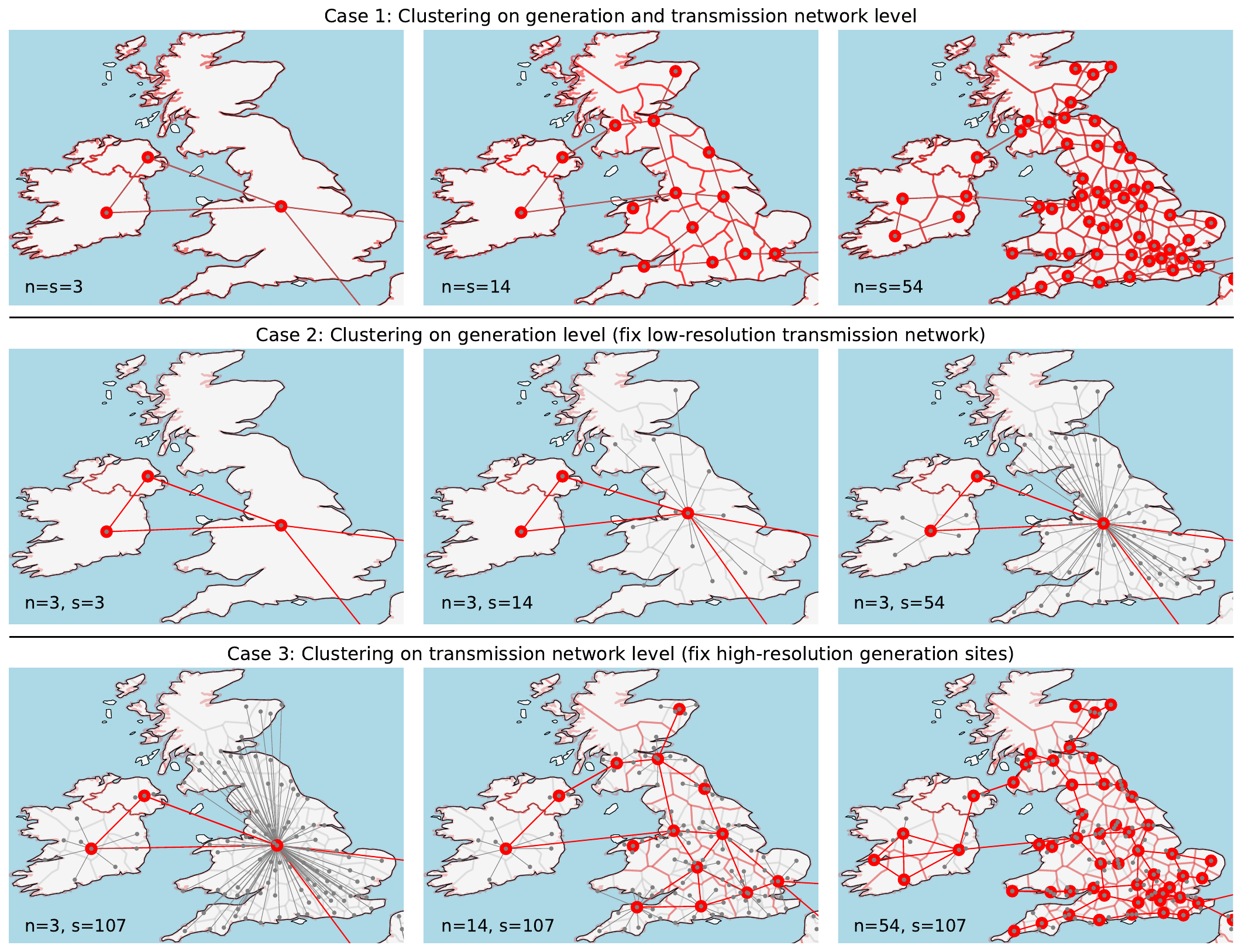}
	\includegraphics[trim=0 0 .3cm 0, clip, width=\textwidth]{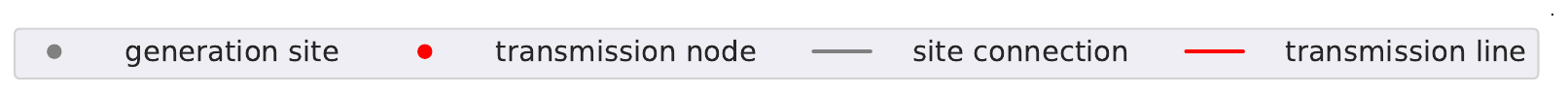}
	\caption{Clustering of network nodes (red, number $n$) and renewable sites (grey, number $s$) in each of the cases (rows) for Ireland and the United Kingdom at different levels of clustering (columns).}\label{fig:clustered-methodology}
\end{figure*}

%https://tex.stackexchange.com/questions/98388/how-to-make-table-with-rotated-table-headers-in-latex
%Dynamic analysis = Small-angle stability, RMS, EMT
\begin{table*}[t]
	\centering
	\setlength{\tabcolsep}{3pt}
	\begin{tabular}{@{} llp{10cm} @{}}
		\toprule
		{\bf Case }   & {\bf Short name} & {\bf Description}  \\
		\midrule
		1 &  Simultaneous clustering &  Successive increase in number of generation sites $s$ and transmission nodes $n$: $s=n\in \mathcal{B}$ \\
		2 &  Clustering on siting resolution & Fix the transmission network to one-node-per-country-zone $n=37$ and increase the number of generation and storage sites $s\in \mathcal{B}$ \\
		3 &  Clustering on network nodes & Maintain a high resolution of generation sites $s=1024$ and successively increase the number of transmission nodes $n\in \mathcal{B}$ \\
		\bottomrule
	\end{tabular}
	\caption{Case descriptions. ($\mathcal{B} = \{37\} \cup
		\bigl\{\lfloor\sqrt{2^i}\rfloor\bigr\}_{i=11,...,20} = \{37, 45, 64, 90, 128, \dots 1024\}$) }
	\label{tab:cases}
\end{table*}

To separate the effects of the spatial resolution on the renewable
resources and the network, we consider three cases in which they are
clustered differently. The three cases are summarized in Table
\ref{tab:cases} and shown graphically in Figure
\ref{fig:clustered-methodology} for each case (rows) and for each
level of clustering (columns).

In {\bf Case 1} the wind and solar sites are clustered to the same
resolution as the network. The number of clusters is varied between
37, the number of country-zones, and 1024, which represents the maximum resolution for which
generation, transmission and storage investment can be co-optimized in reasonable
time.  The number of nodes is increased in half-powers of 2, so that
nine different resolutions are considered: $\mathcal{B} = \{37\} \cup
\bigl\{\lfloor\sqrt{2^i}\rfloor\bigr\}_{i=11,...,20}$.

In {\bf Case 2} network bottlenecks inside each country-zone are
removed so that there are only 37 transmission nodes, and only the
resolution of the wind and solar generation is varied. Inside each
country-zone, all wind and solar generators are connected to the central
node. This allows the optimization to exploit the best wind and solar
sites available.

Finally in {\bf Case 3} we fix a high resolution of renewable sites and vary
the number of network nodes, in order to explore the effects of network
bottlenecks. Each renewable site is connected to the nearest network
node, where the transmission lines, electricity demand, conventional generators and
storage are also connected.

For each case we optimize investments and operation for wind and solar
power, as well as open cycle gas turbines, batteries, hydrogen storage
and transmission. Flexibility from existing hydroelectric power plants
is also taken into account. The model is run with perfect foresight at
a 3-hourly temporal resolution over a historical year of load and
weather data from 2013, assuming a 95\% reduction in CO$_2$ emissions
compared to 1990. The temporal resolution is 3-hourly to capture
changes in solar generation and electricity demand while allowing reasonable
computation times. The technology selection is also limited for
computational reasons. More details on the investment optimization can be found in subsection \ref{sec:investmentoptimisation}.

For each simulation we also vary the amount of new transmission that
can be built, in order to understand the effect of possible grid
reinforcements on the results. The model is allowed to optimize new
transmission reinforcements to the grid as it was in 2018, up to a
limit on the sum over new capacity multiplied by line length measured
relative to the grid capacity in 2018.  For example, a transmission
expansion of 25\% means that on top of 2018's grid, new lines
corresponding to a quarter of 2018's grid can be added to the network.
The exact constraint is given in equation \eqref{eq:linecap} in subsection \ref{sec:investmentoptimisation}.

\subsection{Network preparation}

Before the clustering algorithm can be applied to the network, several simplifications are applied to the data.

In order to avoid the difficulty of keeping track of different voltage levels as the network is clustered, all lines are mapped to their electrical equivalents at 380~kV, the most prevalent voltage in the European transmission system. If the original reactance of the line $\ell_{i,j}$ was $x_{i,j}$ at its original voltage $v_{i,j}$, the new equivalent reactance becomes
\begin{equation}
	x_{i,j}' = x_{i,j} \left(\frac{380 \textrm{ kV}}{v_{i,j}}\right)^2.
\end{equation}
This guarantees that the per unit reactance is preserved after the equivalencing.

The impedances and thermal ratings of all transformers are neglected, since they are small and cannot be consistently included with the mapping of all voltage levels to 380~kV.

\iffalse

First, we consider $DC$ lines, referred to as links, of the network and the nodes they connect. Intermediate nodes without additional connections are aggregated. We will refer to such sequences as "stubs". The following visualisation of a stub depicts nodes with an indexed dot $\cdot_\mathrm{o}$:
\begin{equation*}
	\cdot_0 - \cdot_1 - ... - \cdot_n \quad \mathrm{becomes} \quad \cdot_0 - \cdot_n \,.
\end{equation*}
All the stub-nodes are to one representative node$_c$. The set of nodes to be aggregated is determined by
\begin{equation*}
	N_{c} = \{\mathrm{node}_i:\ \|x_i - x_c\|_2\ \mathrm{is\ minimal}\}\,,
\end{equation*}
where $x\in \mathbb{R}^2$ denotes the geographical location of the node and $x_c$ is the first (node$_0$) or last (node$_n$) node in a stub sequence. However, a stub can also be an isolated $DC$ cycle - that is, if node$_n$ again connects to node$_0$ by a link. In this case, node$_0$ and node$_n$ are essentially arbitrary neighbours.

Inter-cluster links, i.e. links connecting nodes in $N_c$ with nodes in $N_d$, given by the set $N_{c,d}$
\begin{equation}\label{eq:linkmap}
	N_{c,d} = \{(\mathrm{link}_i, \mathrm{link}_j): A_{ij}=1,\ \mathrm{node_i} \in N_c,\ \mathrm{node}_j\in N_d\}\,,
\end{equation}
are aggregated to one representative. $A$ is the adjacency matrix of the associated undirected network graph $\mathcal{G}$. As links are always DC transmission systems, the aggregated one is also a DC carrier. Attributes of the representative link are given in table 2\ref{tab:links}.

\fi

Univalent nodes, also known as dead-ends, are removed sequentially until no univalent nodes exist. That is, if node $i$ has no other neighbor than node $j$, then node $i$ is merged to node $j$. We repeat the process until each node is multi-valent and update the merged node attributes and its attached assets (loads, generators and storage units) according to the rules in Table \ref{tab:nodes}.

HVDC lines in series or parallel are simplified to a single line $\ell$ using the rules in Tables \ref{tab:linesseries} and \ref{tab:linesparallel}. Capital costs per MW of capacity for HVDC lines $\ell_{i,j}$ with length $l_{\ell_{i,j}}$ and a fraction $u_{\ell_{i,j}} \in [0,1]$ underwater are given by
\begin{align}
	c_{i,j} = 1.25 \cdot l_{i,j} \cdot \left(u_{i,j} \cdot c_{\mathrm{marine}} + (1-u_{i,j})\cdot c_{\mathrm{ground}}\right) \,, \nonumber
\end{align}
where $c_\mathrm{marine}$ is the capital cost for a submarine connection and $c_{\mathrm{ground}}$ for an underground connection. The factor of $1.25$ accounts for indirect routing and height fluctuations.

%Annualised fixed costs at each aggregated node $c$ and storage/generation technology are
%\begin{align*}
%c_{c,s} \mapsto c_{c,s} + \sum_{j\in N_c} \mathrm{Dij}(A^{w(c_l)})_{c,j} \,,
%\end{align*}
%where $A^{w(c_l)}$ is the weighted adjacency matrix with weights corresponding to the costs of each link. Dij($\cdot$) denotes the %Dijkstra algorithm: With the weighted adjacency matrix as an input parameter, it returns the costs of the cheapest path to reach node$_j$ %starting from node$_c$.

\subsection{Clustering methodology}\label{sec:clusteringdetails}

\begin{table}\centering
	\caption{Notation}         \label{tab:notation}
	\begin{tabular}{@{} lp{5cm} @{}}
		\toprule
		symbol & meaning\\
		\midrule
		& general abbreviations \\
		\midrule
		$s$ & technology type \\
		$t$ & time point \\
		$i,j$ & nodes in high resolution network \\
		$c,d$ & clustered nodes \\
		$\ell_{i,j}$ & high resolution line connecting nodes $i$ and $j$\\
		$\ell_{c,d}$ & aggregated representative line connecting clusters $c$ and $d$\\
		$N_{c}$ & set of high resolution nodes in cluster $c$ \\
		$N_{c,d}$ & set of high resolution lines between clusters $c$ and $d$ \\
		$RE$ & set of renewable generator technologies \\
		$CG$ & set of conventional generator and storage technologies \\
		\midrule
		& line attributes \\
		\midrule
		$x_{i,j}$ & reactance of line $\ell_{i,j}$ \\
		$v_{i,j}$ & voltage of line $\ell_{i,j}$ \\
		$c_{i,j}$ & capital costs for line $\ell_{i,j}$\\
		$l_{i,j}$ & length of line $\ell_{i,j}$ \\
		$F_{i,j}$ & capacity of line $\ell_{i,j}$\\
		$f_{i,j,t}$ & flow of line $\ell_{i,j}$ at time $t$ \\
		$c_{\substack{\mathrm{marine}/ \mathrm{ground}}}$ & capital costs for a submarine/ underground connection \\
		\midrule
		& nodal and technology attributes \\
		\midrule
		$x_i$ & coordinates of node $i$ in $\mathbb{R}^2$ \\
		$w_i$ & nodal weighting \\
		$e_s$ & CO$_2$ emissions of technology $s$ \\
		$w_{i,s}$ & nodal technology weighting \\
		$c_{i,s}$ & annualised fixed costs \\
		$G_{i,s}$ & (optimal) capacity of technology $s$ at node $i$\\
		$G^\mathrm{max}_{i,s}$ & maximal installable capacity of technology $s$ at node $i$\\
		$o_{i,s}$ & variable costs of technology $s$ at node $i$ \\
		$E_{i,s}$ & storage energy efficiency \\
		$\eta_{i,s}$ & storage losses or efficiencies at node $i$ for technology $s$\\
		$w_t$ & time weighting \\
		$d_{i,t}$ & demand per node $i$ and time $t$ \\
		$\bar{g}_{i,s,t}$ & capacity factor for RE $\in [0,1]$\\
		$g_{i,s,t}$ & dispatch in node $i$ of technology $s$ at time $t$\\
		$e_{i,s,t}$ & energy level of technology $s$ in node $i$  at time $t$\\
		\midrule
		& graph related attributes \\
		\midrule
		$K_{i,\ell}$ & incidence matrix \\
		$C_{\ell,c}$ & Cycle matrix, here, $c$ represents a cylce\\
		\bottomrule
	\end{tabular}
\end{table}

Different methods have been used to cluster networks in the
literature. We chose a version of $k$-means
clustering \cite{Hartigan1979} based on the geographical location of
the original substations in the network, weighted by the average load
and conventional capacity at the substations, since this represents
how the topology of the network was historically planned to connect
major generators to major loads. It leaves the long transmission lines between regions, which are expensive to upgrade and are more likely to encounter low local acceptance, unaggregated, so that these lines can be optimized in the model. Regions with a high density of nodes, for example around cities, are aggregated together, since the short lines between these nodes are inexpensive to upgrade and rarely present bottlenecks.
Geographical $k$-means clustering has the advantage over other clustering methods of not making any assumptions about the future generation, storage and network capacity expansion.

Other clustering methods applied in the literature are not suitable for the co-optimization of supply and grid technologies: these include clustering based on electrical distance using $k$-medoids \cite{Blumsack2009, CotillaSanchez}, a modified version of $k$-medoids to avoid assigning both end nodes
of a critical branch to the same zone \cite{eHighways2050}, hierarchical clustering \cite{BIENER2020106349}, or $k$-decomposition and eigenvector partitioning
\cite{TEMRAZ1994301} (which
we do not use because we want to optimize new grid reinforcements that
alter electrical distances), spectral partitioning of the graph Laplacian
matrix \cite{Hamon2015} (avoided for same reason), an adaptation of
$k$-means called $k$-means$++$ combined with a max-$p$ regions algorithm
applied to aggregate contiguous sites with similar wind, solar and
electricity demand \cite{Siala2019} (avoided since we want a coherent clustering of
all network nodes and assets), hierarchical clustering based on a database of electricity demand, conventional generation and renewable profiles including a synthesized grid \cite{Kueppers2020} (avoided for the same reason and because we do not want to alter the topology of the existing transmission grid), $k$-means clustering based on renewable resources as well as economic, sociodemographic and geographical features \cite{spatialpatterns} (avoided because we need a clustering focused on network reduction), as well as clustering based on zonal Power Transfer Distribution
Factors (PTDFs) to detect congestion zones \cite{Cheng2005}, to yield the same flow patterns as the original network \cite{Oh2010} or to analyse  policy options and emissions \cite{Shi2015} (avoided because they encode electrical parameters that change with reinforcement), Available Tranfer Capacities
(ATCs) \cite{Shayestah2015} (avoided because they depend on pre-defined dispatch patterns) and  locational marginal prices
(LMP) \cite{Singh2005} (again avoided because they depend on pre-defined dispatch patterns).

We do not allow nodes in different countries or different synchronous zones to be clustered together, so that we can still obtain country-specific results and so that all HVDC between synchronous zones are preserved during the aggregation. This results in a minimum number of 37 clustered nodes for the country-zones. First we partition the desired total number $n$ of clusters between the 37 country-zones, then we apply the $k$-means clustering algorithm within each country-zone.

In order to partition the $n$ nodes between the 37 country-zones, the following minimisation problem is solved
\begin{align} \label{eq:distributeclusters}
	\mathrm{argmin}_{\{n_z\}\in\mathbb{N}^{37}}\sum_{z=1}^{37}\left(n_z -  \frac{L_z}{\sum_y L_y} n\right)^2 \,,
\end{align}
where $L_z$ is the total load in each country-zone $z$. An additional constraint ensures that the number of clusters per country-zone matches the desired number of clusters for the whole network: $\sum_z n_z = n$.

Then the $k$-means algorithm is applied to partition the nodes inside each country-zone into $n_z$ clusters. The algorithm finds the partition that minimizes the sum of squared distances from the mean position of each cluster $x_c\in \mathbb{R}^2$ to the positions $x_i\in \mathbb{R}^2$ of its members $i\in N_c$
\begin{align} \label{eq:kmeans}
	\mathrm{min}_{\{x_c\in \mathbb{R}^2\}} \sum_{c=1}^k \sum_{i\in N_c} w_i \cdot \|x_c - x_i\|_2 \,.
\end{align}
Each node is additionally assigned a normalised weighting $w_i$ based on its nominal power for conventional generators and averaged load demand:
\begin{align} \label{eq:busweightings}
	w_{i} &= \frac{\sum\limits_{s_\mathrm{conv.}}G_{i,s}}{\sum\limits_{s_\mathrm{conv.}}\sum_{i=1}^{B} G_{i,s}} + \frac{d_{i,T}}{\sum_{i=1}^{B} d_{i,T}} \,, \quad \forall i
\end{align}
where $d_{i,T}$ corresponds to the averaged demand over the considered time period $T$. $w_i$ is normalised according to $\lfloor \frac{100\cdot w_i}{\|w\|_\mathrm{max}} \rfloor$.

The optimization is run with $n_\mathrm{init} = 10^3$ different centroid seeds, a maximum number of iterations for a single run of $\mathrm{max}_\mathrm{iter} = 3\cdot 10^4$ and a relative tolerance with regards to inertia to declare convergence of $\varepsilon = 10^{-6}$ .

Attributes of the nodes in $N_c$ and their attached assets are aggregated to the clustered node $c$ according to the rules in Table \ref{tab:nodes}.

Lines connecting nodes $N_c$ in cluster $c$ with nodes $N_d$ in cluster $c$, given by the set $N_{c,d}$
\begin{equation}\label{eq:linkmap}
	N_{c,d} = \{\ell_{i,j},\ i \in N_c,\ j\in N_d\}\,, \quad \forall c,d
\end{equation}
are aggregated to a single representative line. The length of the representative line is determined using the haversine formula (which computes the great-circle distance between two points on a sphere) multiplied by a factor of $1.25$ to take indirect  routing into account.
The representative line inherits the attributes of the lines $N_{c,d}$ as described in Table \ref{tab:linesparallel}. If any of the replaced lines in $N_{c,d}$ had the attribute that their capacity was extendable, then the aggregated line inherits this extendability.

An analysis of the effects of clustering
on the network flows can be found in the Appendix, Section \ref{sec:flows}.

For Case 1, generators are clustered to the same resolution as the network. Times series containing hourly resolved capacity factors $\bar{g}_{i,s,t}\in [0,1]$ for variable renewable generation are aggregated using a weighted average
\begin{align} \label{eq:capacityaggregation}
	\bar{g}_{c,s,t} = \frac{1}{\sum_{i\in N_c} w_{i,s}} \sum_{i\in N_c} w_{i,s} \cdot \bar{g}_{i,s,t}\,, \quad \forall c,s,t
\end{align}
The resulting capacity factor $\bar{g}_{c,s,t}$ is in $[0,1]$ by definition. For renewables, the weighting $w_{i,s}$ is proportional to the maximal yearly yield for technology $s$ at node $i$, found by multiplying the maximal installable capacity $G^\mathrm{max}_{i,s}$ with the average capacity factor. In the case of conventional technologies the weightings are distributed equally, i.e $w_{i,s} = 1$. Note that there is no relation between the weightings $w_{i,s}$  and the bus weightings $w_i$ of (\ref{eq:busweightings}).

For Case 2, the network is fixed at 37 nodes, and the wind and solar generators are merged in the aggregation step. Time series for VRE availability are aggregated according to (\ref{eq:capacityaggregation}) to their respective resolution.

For Case 3, the network is clustered, but wind and solar generators are not merged in the aggregation step. Their time series remain fixed at high resolution of 1024 nodes.

\subsection{Investment optimisation} \label{sec:investmentoptimisation}

Investments in generation, storage and transmission are optimized in the PyPSA modelling framework \cite{PyPSA}, which
minimises the total system costs. The objective function is
\begin{align*} %\label{eq:systemcosts}
	\min_{\substack{G_{i,s},\ F_\ell, \\ g_{i,s,t},\ f_{\ell,t}}} \Bigl[ \sum_{i=1}^{B} \sum_{s=1}^{S} \Bigl( c_{i,s} G_{i,s} + \sum_{t=1}^{T} w_t o_{i,s}g_{i,s,t} \Bigr) + \sum_{\ell=1}^{L} c_\ell F_\ell \Bigr] \,,
\end{align*}
consisting of the annualised fixed costs $c_{i,s}$ for capacities $G_{i,s}$ at each node $i$ and storage/generation technology $s$, the dispatch $g_{i,s,t}$ of the unit at time $t$ and associated variable costs $o_{i,s}$ multiplied by a weight factor $w_t$ corresponding to the temporal resolution of the system, and the line capacities $F_\ell$ for each line $\ell$ including both high voltage alternating current and direct current lines and their annualised fixed costs $c_\ell$. The time period $T$ runs over a full year at a 3-hourly resolution, so each time period $t$ is weighted with $w_t = 3$.

Investment cost assumptions are provided in Table \ref{tab:costs}, based on projections for the year 2030. Assumptions are based on \cite{dea2019} for wind technologies, \cite{schroeder2013} in case of OCGT, PHS, hydro, run-of-river, \cite{budischak2013} for storage technologies and \cite{etip} for solar technologies. 2030 is chosen for the cost projections since this is the earliest possible time that such a system transformation might be feasible, and because it results in conservative cost assumptions compared to projections for a later date. The only CO$_2$-emitting generators are the open cycle gas turbines with natural gas with specific emissions 0.187 tCO$_2$/MWh$_\textrm{th}$ and fuel cost 21.6 €/MWh$_\textrm{th}$. Investment costs are annualized with a discount rate of 7\%. Lifetimes, efficiencies and operation and maintenance costs can be found in the GitHub repository \cite{PyPSAEurGitHub}.

\begin{table}\centering
	\caption{Technology investment costs with $1\$ = 0.7532$€.}         \label{tab:costs}
	\begin{tabular}{@{} lrl @{}}
		\toprule
		asset & cost & unit \\
		\midrule
		onshore wind  & 1110 & €/kW \\
		offshore wind & 1640 & €/kW \\
		(AC/DC grid connection separate) & & \\
		solar PV utility & 425 & €/kW \\
		solar PV rooftop & 725 & €/kW \\
		open cycle gas turbine & 400 & €/kW \\
		run of river & 3000 & €/kW\\
		\midrule
		pumped hydro storage & 2000 & €/kW \\
		hydro storage & 2000 & €/kW\\
		battery storage & 192 & \$/kWh \\
		battery power conversion & 411 & \$/kW$_{\textrm{el}}$ \\
		hydrogen storage & 11.3 & \$/kWh \\
		hydrogen power conversion & 689 & €/kW$_{\textrm{el}}$ \\
		\midrule
		HVAC overhead transmission & 400 & €/(MWkm) \\
		HVAC underground transmission & 1342 & €/(MWkm)\\
		HVAC subsea transmission & 2685 & €/(MWkm) \\
		HVDC underground transmission & 1000 & €/(MWkm) \\
		HVDC subsea transmission & 2000 & €/(MWkm) \\
		\bottomrule
	\end{tabular}
\end{table}

The dispatch of conventional generators $g_{i,s,t}$ is constrained by
their capacity $G_{i,s}$
\begin{equation}
	0 \leq g_{i,s,t} \leq G_{i,s} \hspace{1cm} \forall\, i,t,s\in CG
\end{equation}

The maximum producible power of renewable generators depends on the
weather conditions, which is expressed as an availability
$\bar{g}_{i,s,t}$ per unit of its capacity:
\begin{equation}
	0 \leq  g_{i,s,t} \leq \bar{g}_{i,s,t} G_{i,s} \hspace{1cm} \forall\, i,t, s \in RE
\end{equation}

The installable renewable capacity $G_{i,s}$ is constrained by land eligibility for placing e.g. wind turbines or solar panels in each node and for each renewable technology. The land restrictions are derived using the Geospatial Land Availability for Energy Systems (GLAES) tool \cite{Ryberg2018} and are always finite for renewable carriers:
\begin{align}
	G_{i,s} &\leq G^\mathrm{max}_{i,s}<\infty \qquad \forall i, s\in RE
\end{align}

There is no capacity constraint for conventional generators:
\begin{align}
	G_{i,s} &< \infty \qquad \forall i, s\in CG
\end{align}

The energy levels $e_{i,s,t}$ of all storage units have to be consistent between all hours and are limited by the storage energy capacity $E_{i,s}$
\begin{align}
	e_{i,s,t}  = & \eta_0^{w_t} e_{i,s,t-1} +  \eta_{1} w_t\left[g_{i,s,t}\right]^- - \eta_{2}^{-1}w_t \left[g_{i,s,t}\right]^+ \nonumber \\
	& + w_tg_{i,s,t}^\textrm{inflow} - w_tg_{i,s,t}^\textrm{spillage} \nonumber \\
	0 & \leq   e_{i,s,t} \leq E_{i,s}   \hspace{1cm} \forall\, i,s,t
\end{align}
Positive and negative parts of a value are denoted as
$[\cdot]^+= \max(\cdot,0)$, $[\cdot]^{-} = -\min(\cdot,0)$.  The
storage units can have a standing loss $\eta_0$, a charging efficiency
$\eta_1$, a discharging efficiency $\eta_2$, inflow (e.g. river inflow
in a reservoir) and spillage. The energy level is assumed to be
cyclic, i.e. $e_{i,s,t=0} = e_{i,s,t=T}$.

CO${}_2$ emissions are limited by a cap $\textrm{CAP}_{CO2}$, implemented using the
specific emissions $e_{s}$ in CO${}_2$-tonne-per-MWh of the fuel $s$
and the efficiency $\eta_{i,s}$ of the generator:
\begin{equation}
	\sum_{i,s,t} \frac{1}{\eta_{i,s}} w_t \cdot g_{i,s,t}\cdot e_{s} \leq  \textrm{CAP}_{CO2} \quad \leftrightarrow \quad \mu_{CO2} \label{eq:co2cap}
\end{equation}
In all simulations this cap was set at a reduction of 95\% of the
electricity sector emissions from 1990.

The (perfectly inelastic) electricity demand $d_{i,t}$ at each node $i$ must be
met at each time $t$ by either local generators and storage or by the
flow $f_{\ell,t}$ from a transmission line $\ell$
\begin{equation}
	\sum_{s} g_{i,s,t} - d_{i,t} = \sum_{\ell} K_{i,\ell} f_{\ell,t} \hspace{1cm} \forall\, i,t
\end{equation}
where $K_{i,\ell}$ is the incidence matrix of the network. This
equation is Kirchhoff's Current Law (KCL) expressed in terms of the
active power.

In the present paper the linear load flow is used, which has been shown to be a good approximation for a well-compensated transmission network \cite{Stot09}, including for simulations using a large-scale European transmission model \cite{Brown2016}. To guarantee the physicality of the network
flows, in addition to KCL, Kirchhoff's Voltage Law (KVL) must be
enforced in each connected network.  KVL states that the voltage
differences around any closed cycle in the network must sum to
zero. If each independent cycle $c$ is expressed as a directed
combination of lines $\ell$ by a matrix $C_{\ell,c}$ then KVL becomes
the constraint
\begin{equation}
	\sum_{\ell} C_{\ell,c} x_{\ell} f_{\ell,t} = 0 \quad  \hspace{1cm} \forall c,t \label{eq:kvl}
\end{equation}
where $x_\ell$ is the series inductive reactance of line $\ell$. It
was found in \cite{HORSCH2018126} that expressing the linear load flow
equations in this way with cycle constraints is computationally more efficient
than angle- or PTDF-based formulations. Note that point-to-point HVDC
lines have no cycles, so there is no constraint on their flow beyond
KCL.

The flows are also constrained by the line capacities $F_{\ell}$
\begin{equation} \label{eq:transmissionconstraint}
	|f_{\ell,t}| \leq b_{B} \cdot F_{\ell} \hspace{1cm} \forall\,\ell,t
\end{equation}
Although the capacities $F_{\ell}$ are subject to optimisation, no new
grid topologies are considered beyond those planned in the TYNDP 2018
\cite{TYNDP2018}.  The factor $b_B = 0.7$ leaves a buffer of 30\% of
the line capacities to account for $n-1$ line outages and reactive
power flows. The choice of 70\% for $b_B$ is standard in the grid
modelling literature
\cite{Stigler2012,Barrios2015,Fuchs2015,Brown2016} and is also the
target fraction of cross-border capacity that should be available for
cross-border trading in the European Union (EU) by 2025, as set in the
2019 EU Electricity Market Regulation \cite{EMR2019}.

Since line capacities $F_{\ell}$ can be continuously expanded to
represent the addition of new circuits, the impedances $x_\ell$ of the
lines would also decrease. In principle this would introduce a
bilinear coupling in equation (\ref{eq:kvl}) between the $x_\ell$ and
the $f_{\ell,t}$. To keep the optimisation problem linear and
therefore computationally fast, $x_\ell$ is left fixed in each
optimisation problem, updated and then the optimisation problem is
run, in up to 4 iterations to ensure convergence, following the
methodology of \cite{Hagspiel14,Neumann2019}.

In order to investigate the effects of transmission expansion, each
line capacity $F_\ell$ can be extended beyond the capacity in 2018,
$F_\ell \geq F^{2018}_\ell$, up to a
a line volume cap $\textrm{CAP}_{\textrm{trans}}$, which is then
varied in different simulations:
\begin{equation}
	\sum_{\ell} l_\ell \cdot (F_{\ell}-F^{2018}_\ell) \leq  \textrm{CAP}_{\textrm{trans}} \quad \leftrightarrow \quad \mu_{\textrm{trans}} \label{eq:lvcap}
\end{equation}
The caps are defined in relation to 2018's line capacities
$F_{\ell}^{2018}$, i.e.
\begin{equation}
	\textrm{CAP}_{\textrm{trans}} =  x \cdot \sum_{\ell} l_\ell \cdot F_{\ell}^{2018} \label{eq:linecap}
\end{equation}
where $x$ is varied between zero and 50\%.

Since there is a cap on the transmission expansion, the line costs
$c_\ell$ can be set to zero. For the results, costs are added after
the simulation based on the assumptions in Table \ref{tab:costs}.

\subsection{Model output data}
The optimised model returns the spatially-resolved capacity for each technology $G_{i,s}$
as well as the amount of transmission
expansion of each included line $F_{\ell}$. Additionally, the results also
provide dispatch time series for each of the generators $g_{i,s,t}$ and
electricity flows $f_{\ell,t}$ for included lines that obey the constraints
described above in subsection \ref{sec:investmentoptimisation}.

\section{Results}

\begin{figure*}
	\includegraphics[width=\textwidth]{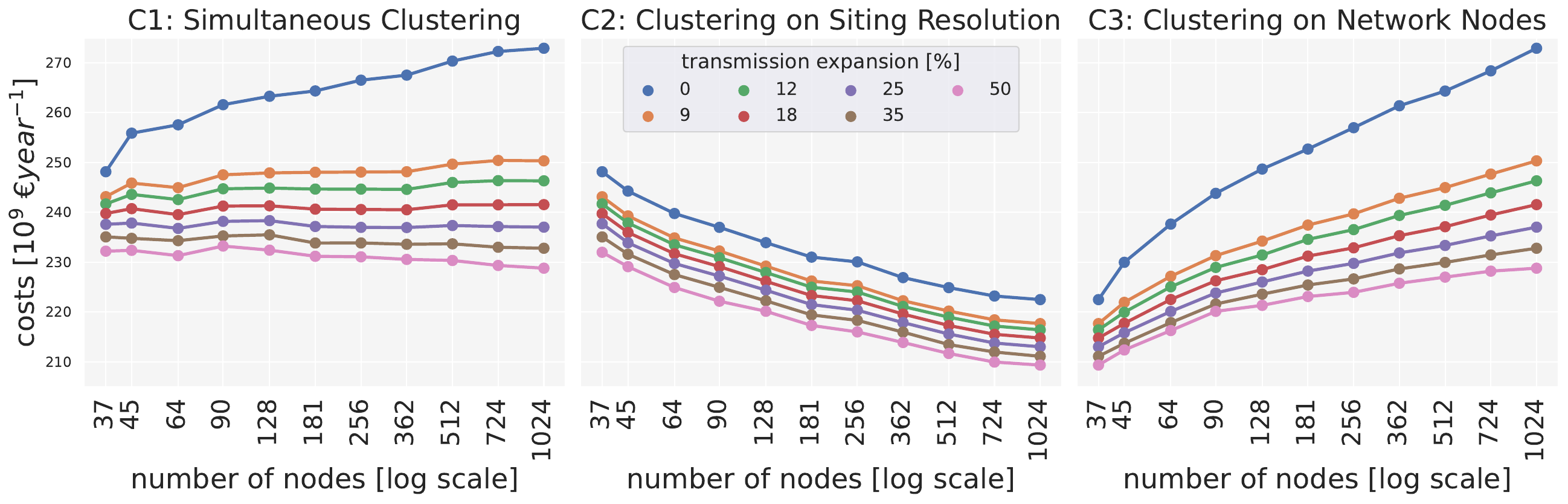}
	\caption{Total annual system costs as a function of the number of clusters for Cases 1, 2 and 3.}
	%Transmission expansion is given as the fraction of today's grid which is added.
	\label{fig:total_costs}
\end{figure*}

\begin{figure*}
	\includegraphics[width=\textwidth]{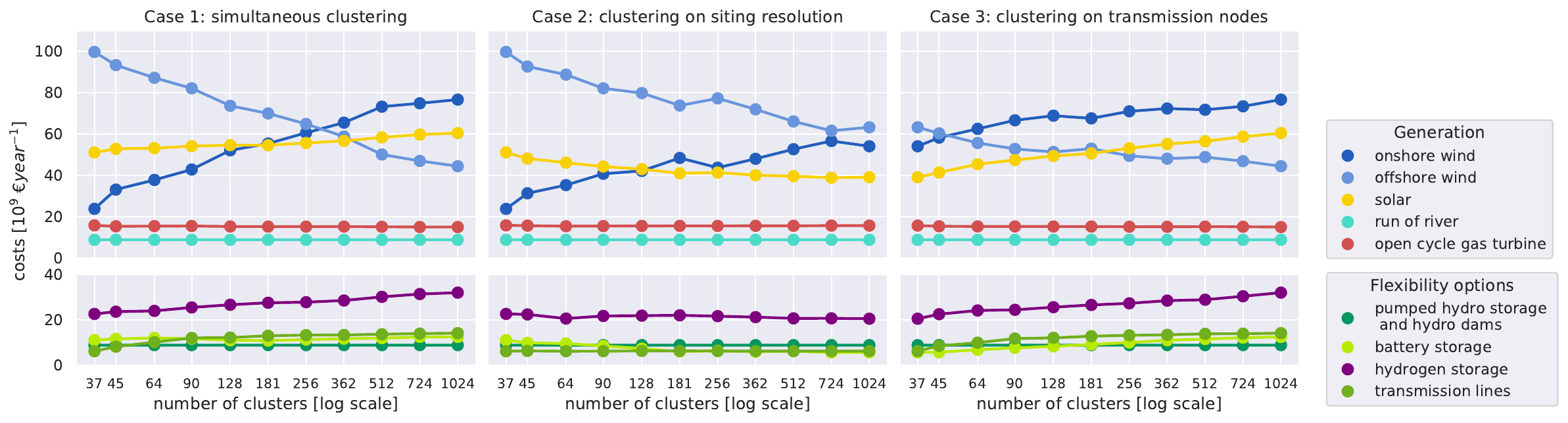}
	\caption{Breakdown of the annual system costs for generation (top) and flexibility options (bottom) as a function of the number of clusters for Cases 1, 2 and 3 when there is no grid expansion.}
	\label{fig:breakdown}
\end{figure*}

\begin{figure} \centering
	\includegraphics[width=8cm]{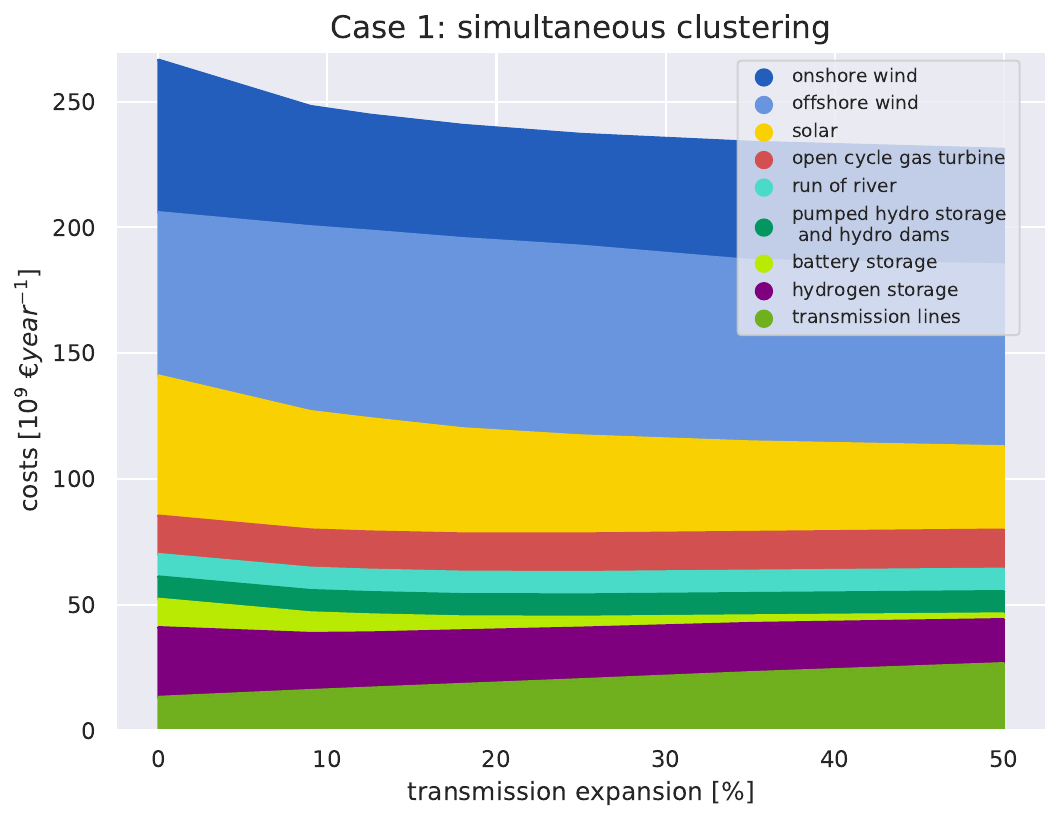}
	\caption{Costs as a function of the transmission expansion level for 256 nodes in Case 1.}
	\label{fig:line_expansion}
\end{figure}

Figure \ref{fig:total_costs} presents the total annual system costs
for each case. To obtain a better understanding of the system
composition, Figure \ref{fig:breakdown} breaks down the total costs
into individual components when there is no grid expansion. In Figure
\ref{fig:line_expansion} we present total system costs for different
grid expansion scenarios for 256 clusters in the simultaneous case
(Case 1). An example map of investments can be found in Figure
\ref{fig:map} for a 25\% grid expansion (a similar level to ENTSO-E's
TYNDP \cite{TYNDP2018}).

\subsection{Case 1 - Increasing number of both generation sites and transmission nodes}

If the resource and network resolutions increase in tandem according
to Case 1 without grid expansion, the total annual system costs in
Figure \ref{fig:total_costs} rise gently with the increasing number of
nodes, reaching a maximum of $273$ billion euros per year at 1024
nodes, which is 10\% more expensive than the solution with 37
nodes. This corresponds to an average system cost of 87~€/MWh.  If
some transmission expansion is allowed, costs are lower, and there is
almost no change in total system costs as the number of nodes is
varied.

However, the fact that costs are flat does not mean that the solutions
are similar: a large shift from offshore wind at low resolution to
onshore wind at high resolution can be observed in the left graph of Figure
\ref{fig:breakdown} (Case 1). This is an indication that spatial resolution can
have a very strong effect on energy modelling results. To understand
what causes this effect, we must examine Cases 2 and 3.

%Investments in other technologies remain similar
%and their locations around Europe also remain stable (see Figure XX).

\subsection{Case 2 - Importance of wind and solar resource granularity}

In Case 2 we use the lowest network resolution of 37 nodes,
corresponding to one-node-per-country-zone, and investigate the effect
of changing the number of wind and solar sites on the results. As the
resolution increases, total costs without grid expansion in Figure
\ref{fig:total_costs} drop by $10$\% from $248$ to $222$ billion euro per year.
Although the slope of the cost curve appears constant, note that the $x$-axis is logarithmic,
so that the rate of cost decrease slows as the number of sites increases.

The cost reduction is driven by strong changes in the investment between
generation technologies, particularly the ratio between offshore and
onshore wind (see Figure \ref{fig:breakdown}). At low spatial
resolution, good and bad onshore sites are mixed together, diluting
onshore capacity factors and making onshore a less attractive
investment. Figure \ref{fig:cfs} in the Appendix shows
how the capacity factors for wind and solar vary across the
continent. While offshore is spatially concentrated and solar capacity
factors are relatively evenly spread in each country-zone, onshore
wind is stronger near coastlines. At high spatial resolution the model
can choose to put onshore wind only at the best sites (within land
restrictions), increasing average capacity factors and thus lower the
per-MWh-cost. (The increasing average capacity factors are plotted in Figure \ref{fig:tech_cap_factors} in the Appendix.) As a result, onshore wind investments more than double from
$24$ to $54$ billion euros per year, while offshore investments drop $37$\% from $100$ to $64$ billion per year and solar by $23\%$. The biggest effect on the technology mix is when going from 37 to around $181$ clusters; beyond that the changes are smaller.

\subsection{Case 3 - Impact of transmission bottlenecks}

In Case 3 we fix a high resolution of wind and solar generators (1024
sites) and vary the resolution of the transmission network to gauge the
impact of transmission bottlenecks. With 37 network nodes many
bottlenecks are not visible, so costs are lower, but as the resolution
increases to 1024 nodes it drives up the costs by $23$\%. Note that because the $x$-axis is logarithmic,
the highest rate of cost increase is when the number of nodes is small.

As can be seen from the breakdown in Figure \ref{fig:breakdown}, the
rising transmission investments from the higher resolution only have a
small contribution to the result. Instead, rising costs are driven
by generation and storage. Unlike Cases 1 and 2, the ratio between the
generation technologies does not change dramatically with the number of
clusters, but the capacities for onshore wind, solar, batteries and
hydrogen storage all rise.

The transmission bottlenecks limit the transfer of power from the best
sites to the load, forcing the model to build onshore wind and solar
more locally at sites with lower capacity factors. Average capacity
factors of onshore wind and solar sink by $11$\% and $6$\% respectively
with no grid expansion (see Figure \ref{fig:tech_cap_factors} in the
Appendix), meaning that more capacity is needed for the
same energy yield.  Curtailment is generally low in the optimal solution (around
3\% of available wind and solar energy) and has less of an effect on
costs (see Figure \ref{fig:curtailment} in the Appendix).

Investment in battery and hydrogen storage rises with the number of
network nodes since the storage is used to balance local wind and
solar variations in order to avoid overloading the grid bottlenecks.

\subsection{Comparison of the three cases}

Separating the effects of resource resolution from network resolution
reveals that the apparent stability of total system costs in Case 1 in
Figure \ref{fig:total_costs} as the number of clusters changes, as
reported in \cite{Hoersch2017}, is deceptive. In fact, the sinking
costs from the higher resource resolution are counter-acted by the
rising costs from network bottlenecks. With no grid
expansion, the system cost of network bottlenecks is double the
benefit of the higher resource resolution.

While these two effects offset each other at the level of total system
costs, they have very different effects on the technology mix.
Resource resolution leads to much stronger investment in onshore wind,
once good sites are revealed. Network bottlenecks have only a weak
effect on the ratio of generation technologies, but lead to lower average capacity factors and drive up storage requirements.

\begin{figure} \centering
	\includegraphics[width=8cm]{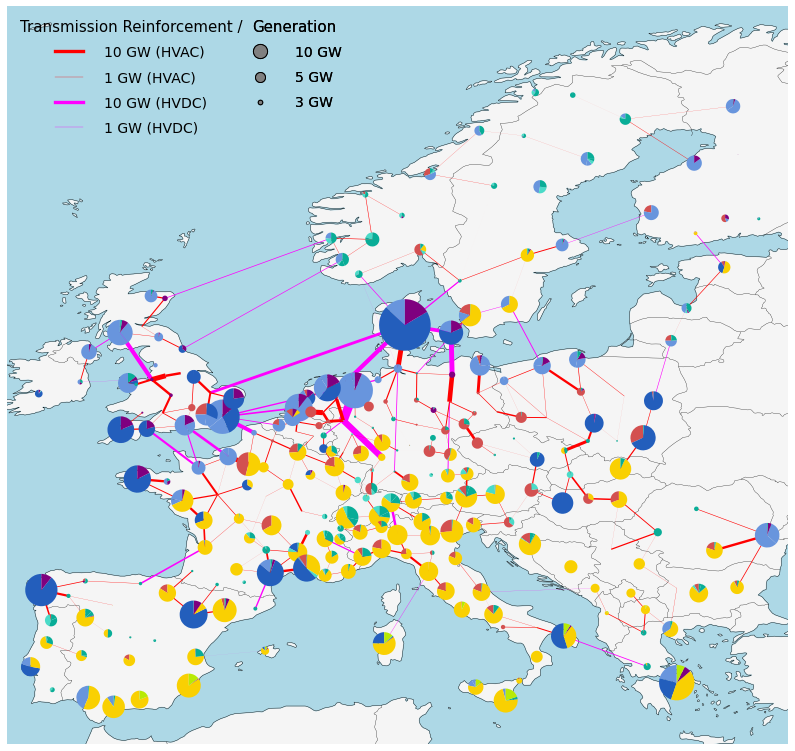}
	\caption{Example of investments with 25\% grid expansion and  256 nodes in Case 1.}
	\label{fig:map}
\end{figure}

\subsection{Benefits of grid expansion}

Grid expansion does not affect the main qualitative features of the
different Cases, but it does have the overall effect of lowering total
system costs.  In Case 1, the total cost-benefit of grid expansion
is highest at around 16\% for a 50\% increase in grid capacity, with the marginal benefit still increasing,
but it is subject to diminishing returns (see Appendix Figure \ref{fig:shadow} for a comparison of the marginal benefit to the cost of transmission). The first 9\% of additional
grid capacity brings total cost savings of up to 8\%, but for each
extra increment of grid expansion, the benefit is weaker.  There is
more benefit from grid expansion at a higher number of nodes, since
the higher network resolution reveals more critical bottlenecks in the
transmission system.

The total savings from 25\% and 50\% grid expansion are around 36 and
44 billion euros per year respectively. In a 2018 study ENTSO-E
examined scenarios with up to 75\% renewable electricity in Europe in
2040 with and without planned TYNDP grid expansions (corresponding to
around 25\% grid expansion), given fixed demand and a fixed generation
fleet. They found that the grid reinforcements reduce generation costs by
43 billion euros per year. This is higher than our cost-benefit for 25\% grid expansion, despite
their study's lower level of renewable electricity, because in our
simulations the generation and storage fleet can be
re-optimised to accommodate the lower level of grid capacity, and because we subtract the costs of new grid reinforcement from the cost-benefit (a contribution of around 3.5 billion euros per year).

%(always ~$1.5$\% savings between two successive reinforcements (i.e 9% and 12.5%) at 512 nodes)
The breakdown of system cost as the grid is
expanded for a fixed number of clusters (256), plotted in Figure
\ref{fig:line_expansion}, reveals how costs are reduced. Although the investment in transmission
lines rises, generation and storage costs reduce faster as investment
shifts from solar and onshore wind to offshore wind. Offshore wind
reduces costs because of its high capacity factors and more regular
generation pattern in time. It can be transported around
the continent more easily with more transmission, and benefits from
the smoothing effects over a large, continental area that grid expansion
enables. The map of investments in Figure \ref{fig:map} shows how
offshore wind is balanced by new transmission around the North Sea,
which smooths out weather systems that roll across the
continent from the Atlantic. Further transmission reinforcements
bring energy inland from the coastlines to load centers. With more
transmission, there is less investment in battery and hydrogen
storage, as a result of the better balancing of weather-driven
variability in space.

Turning to Case 3, we see that grid expansion mitigates the effect of network resolution by allowing
bottlenecks to be alleviated. For a 50\% increase in transmission
capacity, total costs rise by only 4\% from 90 nodes up to 1024 nodes.
%227/220 = 1.0318
The
distribution of investments between technologies also barely changes in this range (see Appendix Figure \ref{fig:breakdown_detail}). This means that a grid resolution of
around 90 nodes can give acceptable solutions for grid expansion
scenarios if computational resources are limited, as long as the wind
and solar resolution is high enough (as in Case 2, 181 generation sites would suffice). Without grid expansion, a higher grid resolution is needed to capture the effects of bottlenecks and achieve reliable results.

\subsection{Computation times and memory}

Besides the poor availability of data at high resolution, one of the
main motivations for clustering the network is to reduce the number of
variables and thus the computation time of the optimisation. In
Appendix Figure \ref{fig:solvingtime} the memory and solving time
requirements for each Case are displayed as a function of the number
of clusters. Both memory and solving time become limiting factors
in Cases 1 and 3, with random access memory (RAM) usage peaking at
around 115~GB and solving time at around 6 days for 1024
clusters. Beyond this number of clusters no consistent convergence in
the solutions was seen.

Case 2, where the network resolution is left low and the resource
resolution is increased, shows seven times lower memory consumption and
up to thirteen times faster solving times compared to Cases 1 and 3 for the
same number of clusters. It is therefore the network resolution rather
than the resource resolution that drives up computational requirements, which it does
by introducing many new variables and possible spatial trade-offs into
the optimisation. Since Case 2 proved relatively reliable for
estimating the ratio between technologies, if not their total
capacity, it may prove attractive to increase the resource resolution rather than
the network resolution if computational resources are limited.

\subsection{Further results}

Further results on curtailment, average capacity factors, the
distribution of technologies between countries, maps, network flows
and shadow prices can be found in the Appendix, as well as a discussion of the
limitations of the model.

\section{Conclusion}

From these investigations we can draw several conclusions. Modellers
need to take account of spatial resolution, since it can have a strong
effect on modelling results. In our co-optimization of generation,
storage and network capacities, higher network resolution can drive up
total system costs by as much as 23\%. Higher costs are driven by the
network bottlenecks revealed at higher resolution that limit access to
wind and solar sites with high capacity factors. On the other hand, resource
resolution affects the balance of technologies by revealing more
advantageous onshore wind sites. In both cases the system costs are
driven more by the useable generation resources than investments in
the grid or storage.

If grid expansion can be assumed, a grid resolution of 90 nodes for
Europe is sufficient to capture costs and technology investments as
long as the solar and onshore wind resolution is at least around 181
nodes.  If grid expansion is not possible, a higher spatial resolution
for the grid is required for reliable results on technology choices. Since grid
expansion is likely to be limited in the future by low public
acceptance, more attention will have to be paid to the computational
challenge of optimizing investments at high spatial granularity.

\section{Data availability}

\subsection{Lead contact}

Please contact the Lead Contact,  Martha M. Frysztacki (martha.frysztacki@kit.edu), for
information related to the data and code described in the following Material and Methods section.

\subsection{Materials Availability}

No materials were used in this study.

\subsection{Data and Code Availability}

All the code and input data from PyPSA-Eur are
openly available online on GitHub and Zenodo \cite{PyPSAEurGitHub,PyPSAEurZenodo010}.
All model output data is available on Zenodo under a Creative Commons Attribution Licence \cite{ZenodoClustering}.

\section{Glossary}

All notation is listed in Table \ref{tab:notation}.

%\subsection{Summary} We summarise that the network resolution mostly plays a role for investment decisions in renewable energy carriers that are subject to weather variability and uncertainty. As storage can balance these insecurities, the spatial resolution also impacts investment decisions in storage options, but affects mostly short-term battery storage. For hydrogen or pumped hydro storage the effect becomes minor. Conventional generation is almost not affected at all because of the carbon dioxide cap of $95\%$ compared to 1990s level, and run of river power generation is also mostly steady as it is not governed by varying availability.

%The effect of generation and transmission, considered jointly for spacial data reduction, does not impact the total system costs because the effect of exploiting good generation sites balances the effect of transmission bottlenecks. But it has a high impact on the composition of renewable carriers.

\section{Acknowledgements}

We thank Martin Greiner, Fabian Neumann, Lina Reichenberg, Mirko Schäfer, David Schlachtberger, Kais Siala and Lisa Zeyen for helpful discussions,
suggestions and comments.  MF, JH and TB acknowledge funding from the
Helmholtz Association under grant no.~VH-NG-1352. The responsibility for the contents lies with the
authors.

\section{Declaration of Interests}

The authors declare that they have no
competing financial interests.

\appendix
\section{Appendix}

\subsection{Preservation of flow patterns with clustering}\label{sec:flows}

\begin{figure} \centering
	\includegraphics[width=9cm]{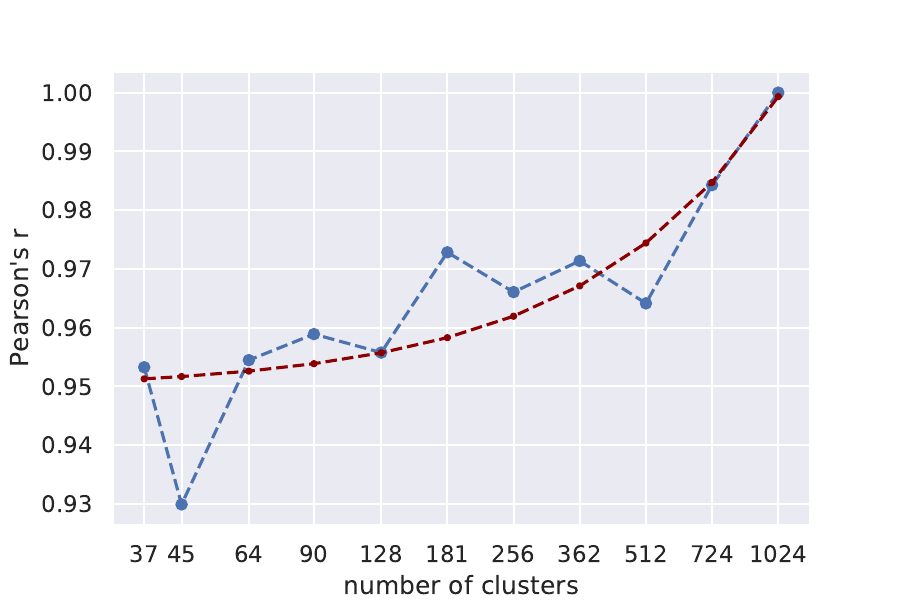}
	\caption{Pearson's correlation coefficient of mapped flows (blue). Note that the x-axis is non-linear, therefore we mark a linear fit to the data (red).} \label{fig:pearson}
\end{figure}

\begin{figure} \centering
	\includegraphics[height=4cm]{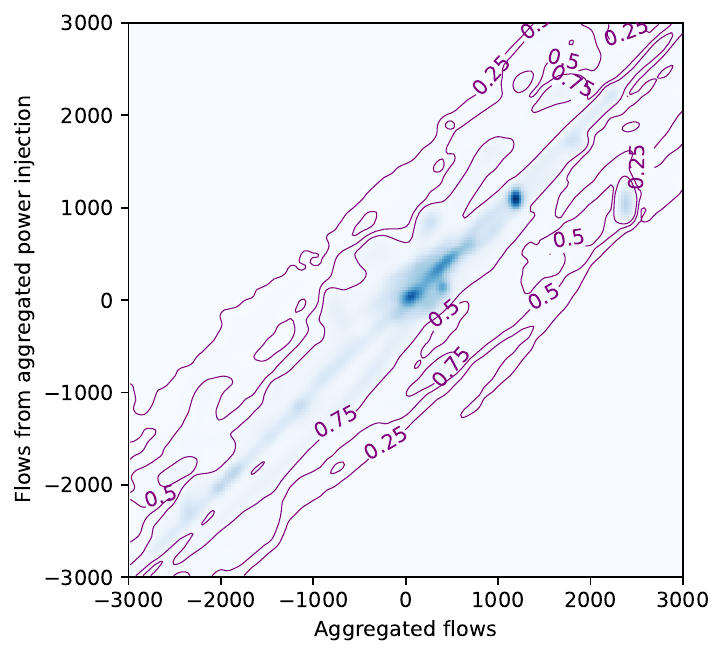}
	\includegraphics[trim=2cm 0 0 0, clip, height=4.cm]{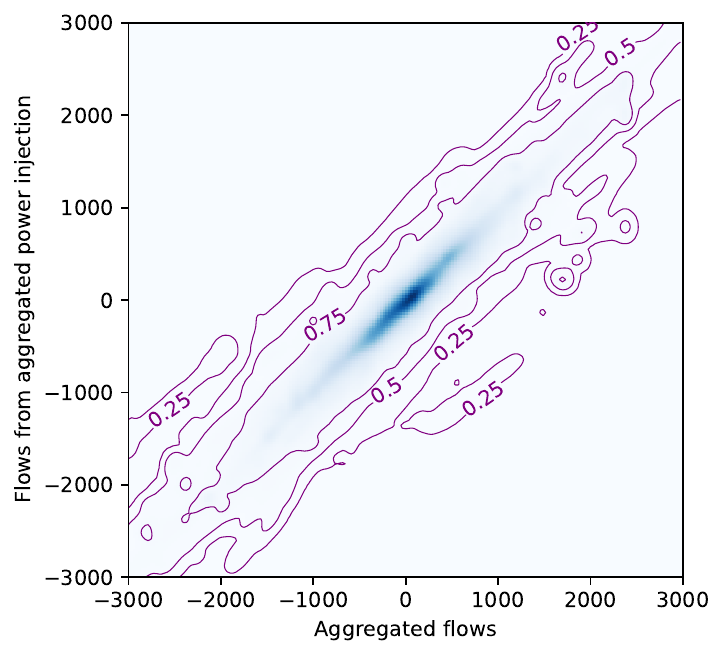}
	\caption{Kernel Density Estimation (KDE) of aggregated flows from a high resolution network grid with 1024 nodes on the $x$-axis and a low resolution grid with 45 nodes (left) and 362 nodes (right) on the $y$-axis. 0.25, 0.5 and 0.75 quantiles of the distribution are displayed as purple isolines around the KDE.} \label{fig:QQ-flowdistribution}
\end{figure}

To understand how well the $k$-means clustering preserves flow patterns,
we took a fixed dispatch pattern for the assets in Europe at high
resolution and examined how the network flows changed as the network
was clustered.

The fixed dispatch was determined by solving the linearised optimal
power flow problem for a 1024-node representation of today's European
electricity system. The asset dispatch was then mapped into the
clustered networks, and a regular linearised power flow was solved in
the clustered network.

If lines $\ell \in N_{c,d}$ in the 1024-node network were mapped to a single
representative line $\ell_{c,d}$ in the clustered network, the
summed flows from the original network $\hat{f}_{c,d,t} = \sum_{\ell \in N_{c,d}} f_{\ell,t}$ (`microscopic flows') were then compared to
the flow $f_{c,d,t}$ in line  $\ell_{c,d}$ of the clustered network (`macroscopic flows').

Figure \ref{fig:pearson} shows the Pearson correlation coefficient between the flows $f_{c,d,t}$ of aggregated lines $\ell_{c,d}$ in the lower resolution network and the summed flows $\hat{f}_{c,d,t}$ of all lines in $N_{c,d}$ in the full resolution network. Red is a linear fit through the points. The distortion from linearity is due to a non-linear scale in the $x$-axis. Even at 37 nodes the correlation between the flows is good (Pearson correlation coefficient above 0.90) and shows
an improving trend until at full 1024-node resolution the flows are once again perfectly equal.

Example density plots of the $\hat{f}_{c,d,t}$ against the $f_{c,d,t}$
for all lines and all times are plotted for different clustering
levels in Figure \ref{fig:QQ-flowdistribution}. The match between the flows is better for higher resolution networks, with a near-diagonal line already for 362 nodes.

For a more probabilistic approach, we perform a kernel density estimation (KDE) by applying a fast Fourier transformation of aggregated flows of the higher resolved network versus the flows of the low resolution network. Aggregated flows $\hat{f}_{c,d,t}$ are considered an estimator for the flow $f_{c,d,t}$ in the representative lower resolution network. The resulting density functions from the KDE are displayed in Figure \ref{fig:QQ-flowdistribution}. For the low resolution network, the probability distribution has two different modes, while a higher resolution network approaches a Gaussian distribution. The variance of the probability density function for a low resolution network is higher than for a high resolution network, as each of the quantile isolines are broader.

\subsection{Maps of capacity factors for wind and solar}
\begin{figure} \centering
	\begin{subfigure}{.55\textwidth} \centering
		\includegraphics[height=7cm]{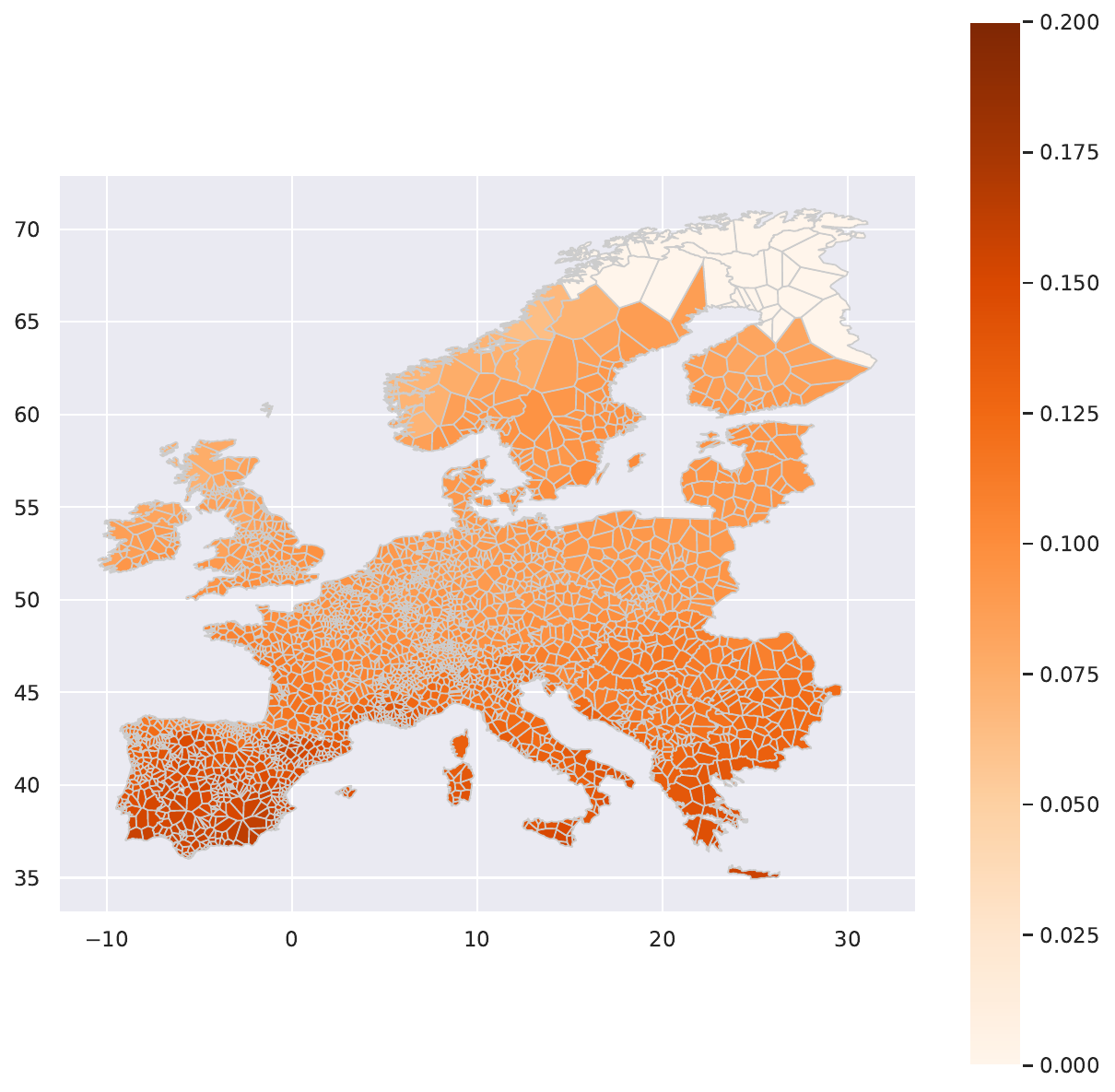}
		\caption{Solar}
		\label{fig:profile_solar}
	\end{subfigure}
	\begin{subfigure}{.55\textwidth} \centering
		\includegraphics[height=7cm]{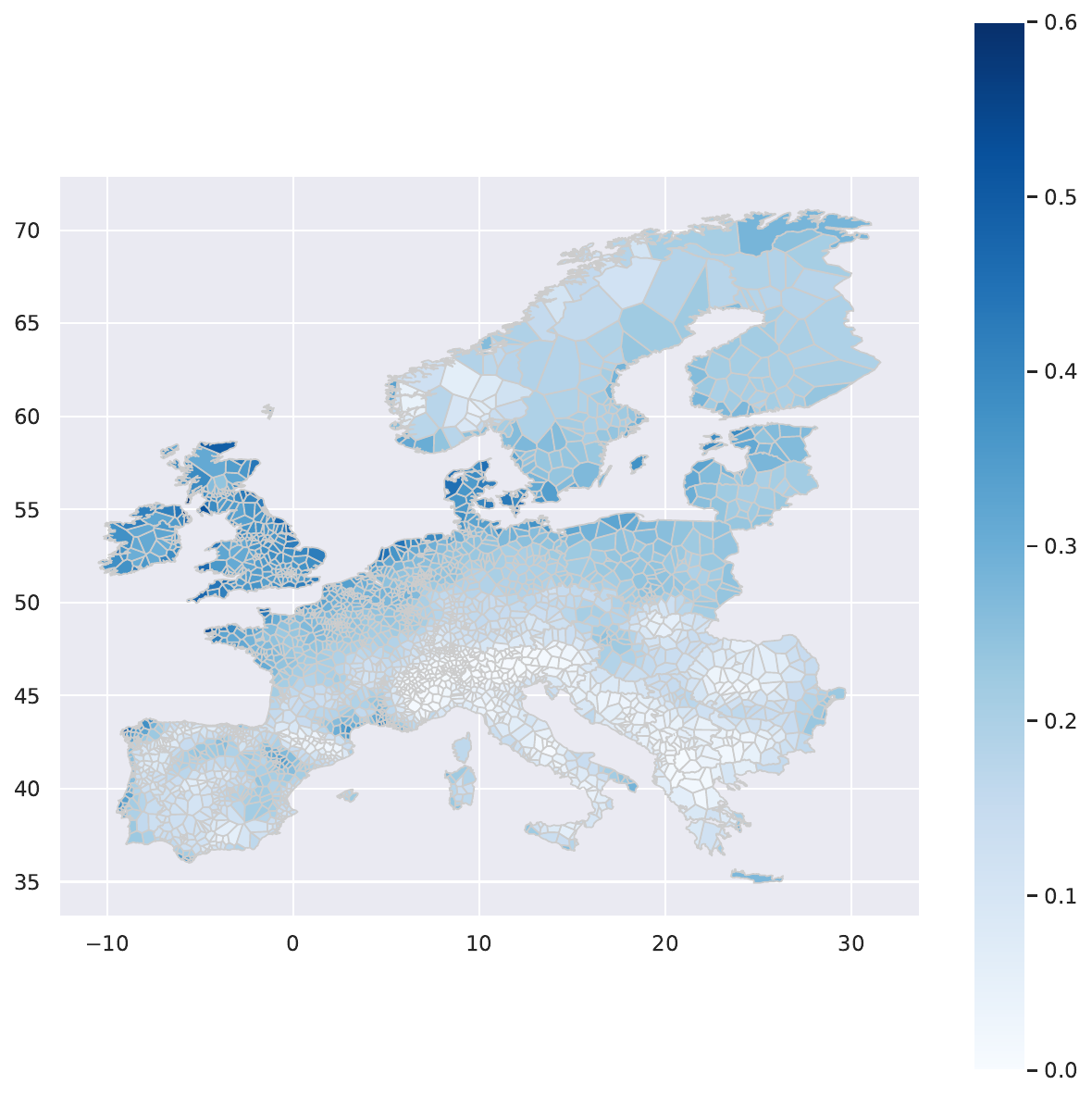}
		\caption{Wind onshore}
		\label{fig:profile_onwind}
	\end{subfigure}
	\begin{subfigure}{.55\textwidth} \centering
		\includegraphics[height=7cm]{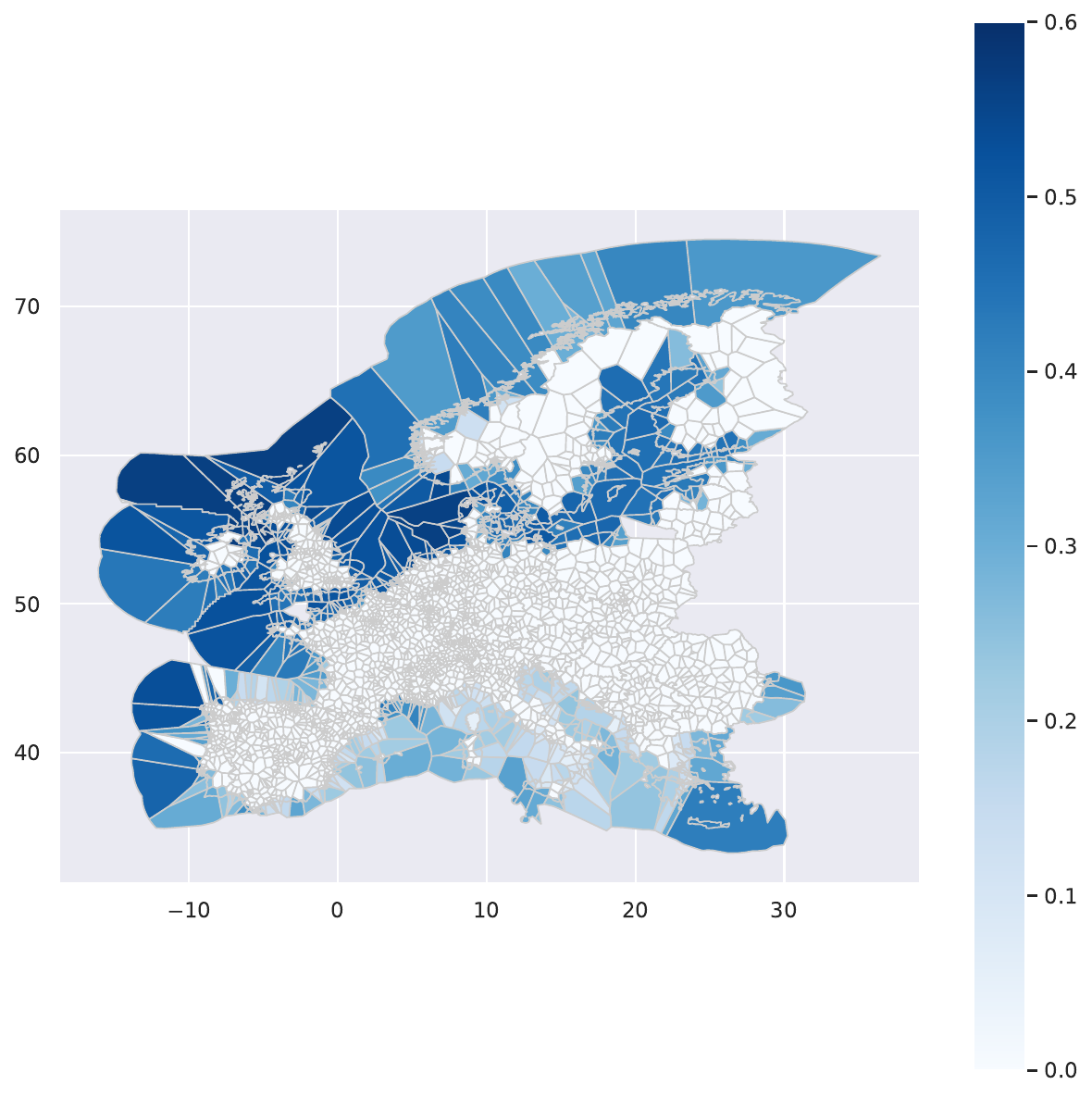}
		\caption{Wind offshore}
		\label{fig:profile_offwind}
	\end{subfigure}
	\caption{Wind and solar capacity factors in Europe for the weather year 2013 at full resolution.}
	\label{fig:cfs}
\end{figure}

Figures \ref{fig:profile_solar}, \ref{fig:profile_onwind}, \ref{fig:profile_offwind} present average capacity factors over the weather year 2013 for solar, wind on- and off-shore respectively, i.e.
\begin{align*}
	\bar{g}_{n,s} = \langle \bar{g}_{n,s,t} \rangle_t \quad \forall n \,,
\end{align*}
where $s\in\{\mathrm{solar},\ \mathrm{wind\ onshore},\ \mathrm{wind\ offshore}\}$. The capacity factors are shown in the Voronoi cells around each of the 1024 node of the original network, i.e. the set of points closest to each node.

The graphics show that capacity factors for solar are decreasing from South to North while those for wind are increasing towards the North and Baltic Sea. The average capacity factors are spatially correlated, but as they are aggregated over larger and larger areas using the weighted average from the clustering approach in equation \eqref{eq:capacityaggregation}, they decline as bad sites are mixed with good sites. This is reflected in Figure \ref{fig:tech_cap_factors}, which shows how the average capacity factors per technology for the generation fleet optimized over the whole of Europe change with the clustering.

\subsection{Breakdowns for multiple transmission expansion scenarios}

\begin{figure*}
	%trim={<left> <lower> <right> <upper>}
	\includegraphics[width=\textwidth]{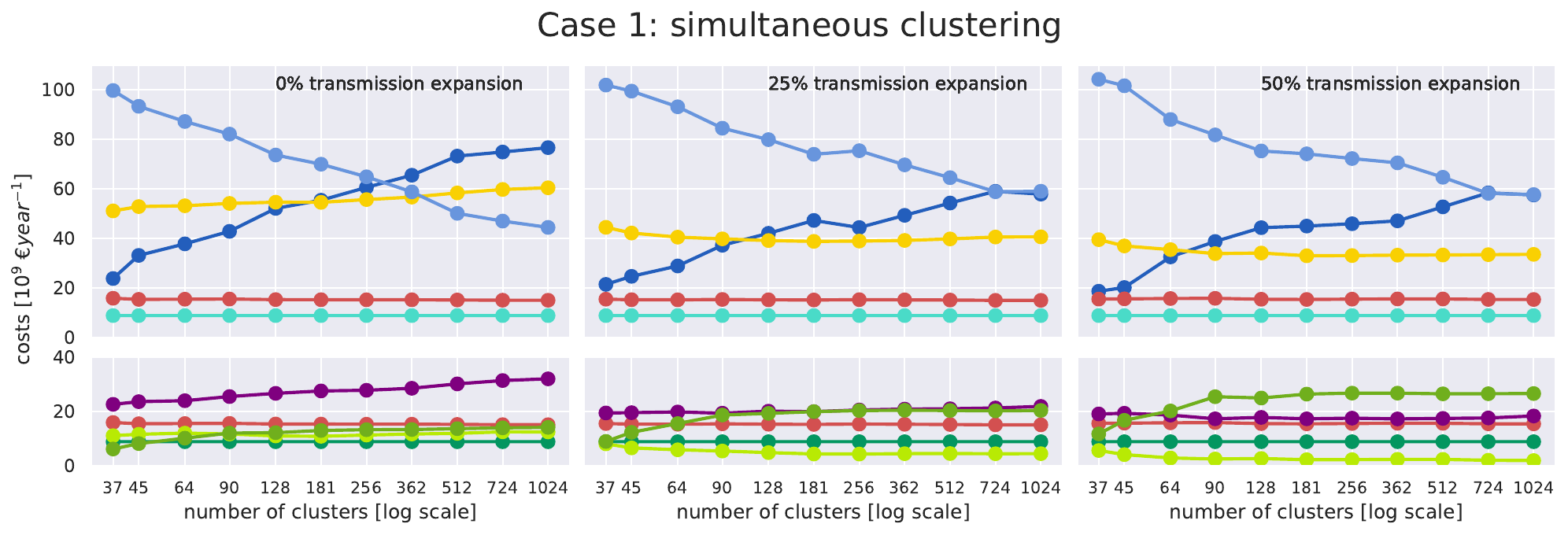}
	\includegraphics[width=\textwidth]{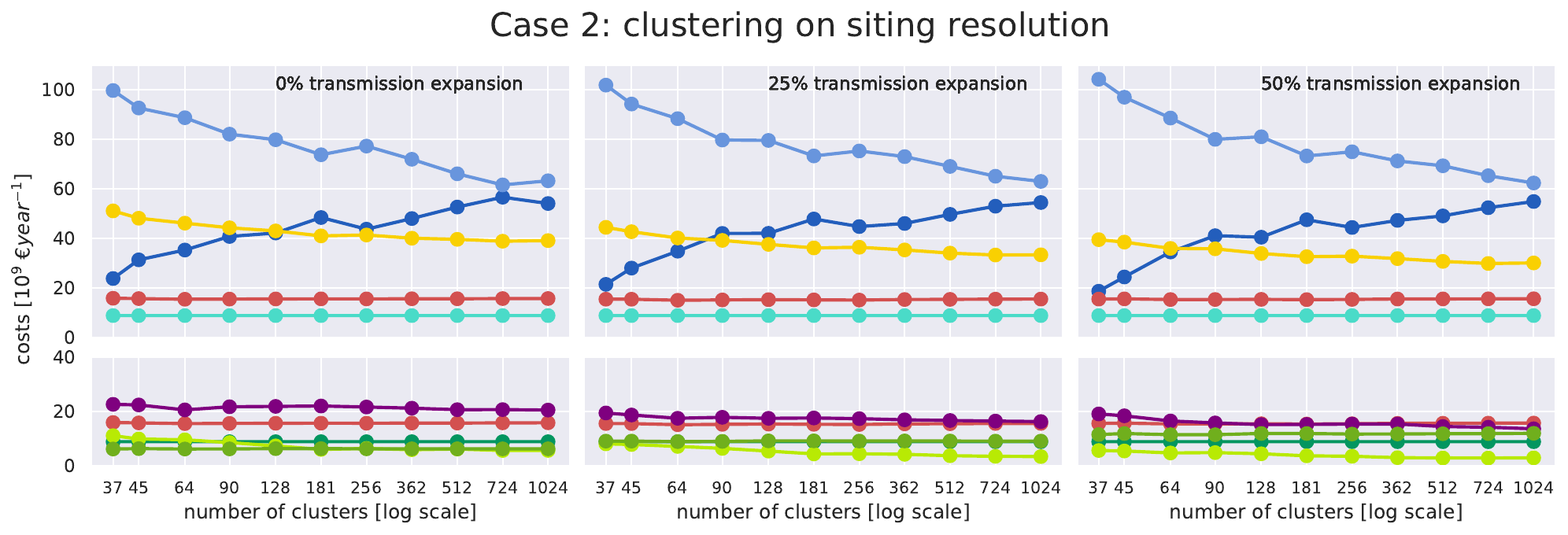}
	\includegraphics[width=\textwidth]{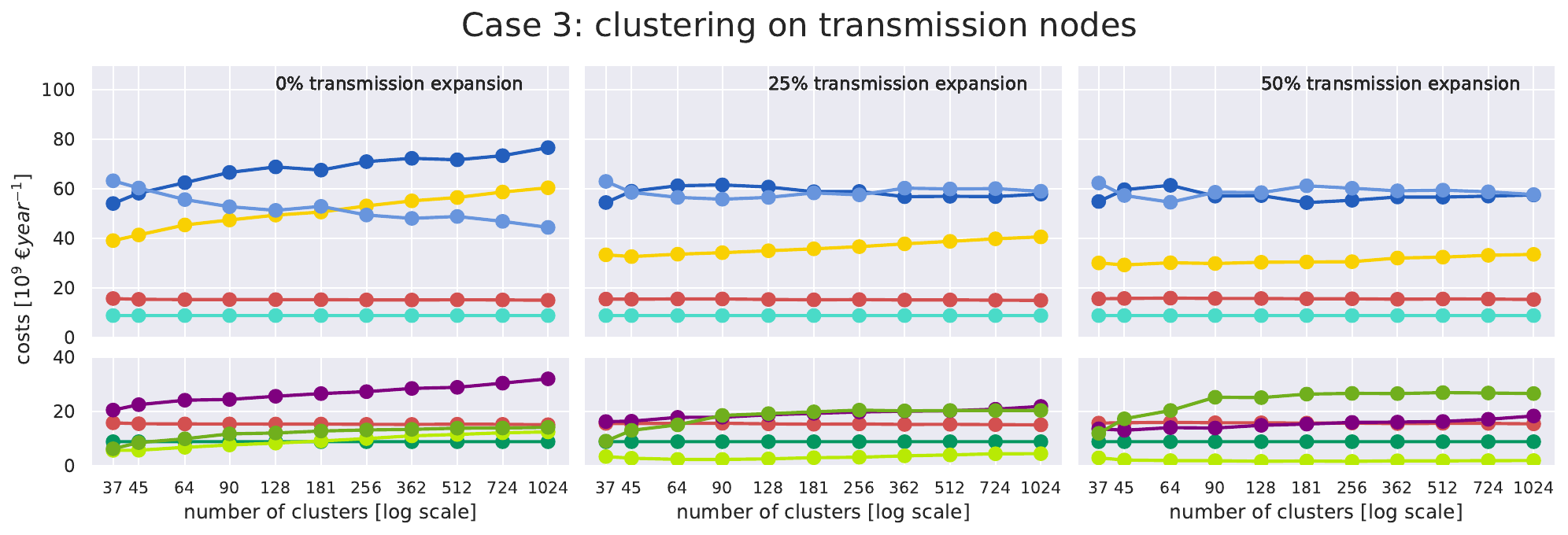}
	\caption{Technology breakdown of the annual system costs for generation (top) and flexibility options (bottom) as a function of the number of clusters for Cases 1, 2 and 3. Cases correspond to the rows, while transmission expansion scenarios correspond to the columns.}
	\label{fig:breakdown_detail}
\end{figure*}

Figure \ref{fig:breakdown_detail} shows an extension of the cost breakdowns in Figure \ref{fig:breakdown} from the scenario with no transmission to scenarios with 25\% and 50\% grid expansion. The general trends are the same as for the scenario without grid expansion, but grid expansion generally allows more wind capacity to be built, resulting in lower investment in solar, batteries and hydrogen storage, as was seen in Figure \ref{fig:line_expansion}.

\subsection{Average capacity factors per technology}

To understand how the model exploits the best available resource sites per node, we examine a time-averaged technology-specific capacity factor $\bar{g}_{s}$. The capacity factor is weighted by how much capacity $G_{n,s}$ of technology $s$ was built at each node $n$ with time-averaged capacity factor $\bar{g}_{n,s}=\langle\bar{g}_{n,s,t} \rangle_t$.
\begin{align*}
	\bar{g}_{s} := \frac{\sum_{n}  \bar{g}_{n,s}\cdot G_{n,s} }{ \sum_n G_{n,s}}\,.
\end{align*}

\begin{figure*} \centering
	\includegraphics[width=\textwidth]{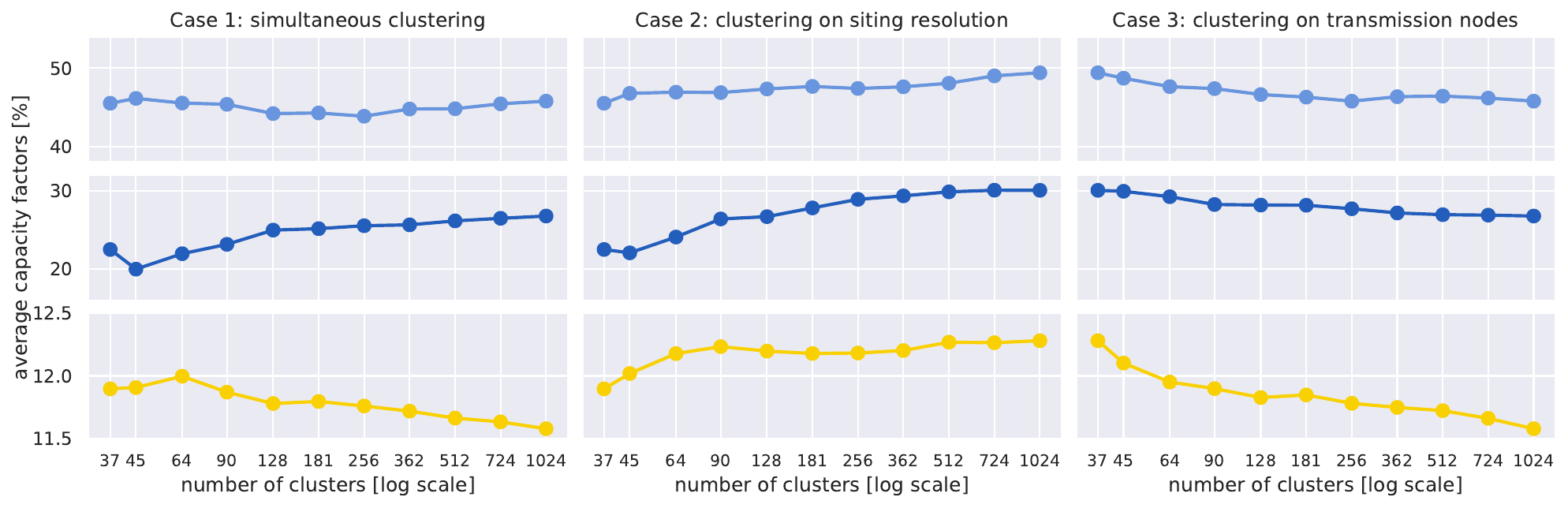}
	\caption{Average capacity factors for each technology for the no transmission expansion scenario in all three cases.} \label{fig:tech_cap_factors}
\end{figure*}

We present this technology-specific capacity factor in Figure \ref{fig:tech_cap_factors} for all three cases with the no-expansion transmission scenario, i.e. where $F_\ell = F_\ell^{2018}$.

As the number of clusters increases, Case 2 has a larger variety of sites per node to choose where capacity should be installed optimally and is not restricted by transmission constraints beyond country-zones. Therefore, the more sites are available, the higher the weighted capacity factor is because it is not mixed with lower capacity factor sites in equation \eqref{eq:capacityaggregation}. The highest resolution of Case 2 is also the lowest resolution of Case 3: many resource sites and only one node per country-zone. As the number of nodes in Case 3 increases while the same sites are available, transmission bottlenecks force the model to build more capacity in locations of worse capacity factors. Therefore, the capacity factors drop again. For Case 1, where resource resolution and network resolution change in tandem, the resource resolution dominates and we see increasing capacity factors like in Case 2.

\subsection{Curtailment per technology}

Curtailment is the amount of energy that is available in theory but cannot be injected into the grid because of transmission constraints or a lack of demand:
\begin{align*}
	\bar{g}_{n,s,t}\cdot G_{n,s}-g_{n,s,t}
\end{align*}

\begin{figure*} \centering
	\includegraphics[width=\textwidth]{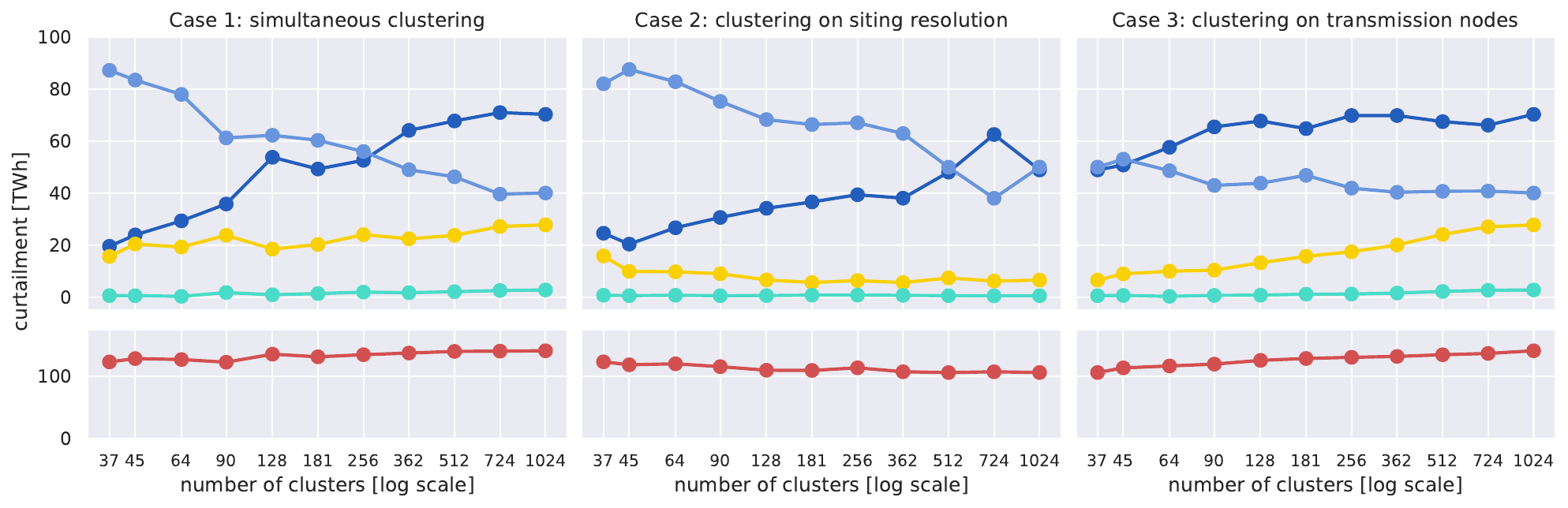}
	\caption{Curtailment for the no transmission expansion scenario in all three cases.} \label{fig:curtailment}
\end{figure*}

Figure \ref{fig:curtailment} shows total curtailment per technology in all Cases. Curtailment in all situations is low (less than $4\%$ of total demand). Curtailment increases with higher network resolution in both the Cases $1$ and $3$ that incorporate transmission constraints, while it is gently decreasing with resource resolution in Case $2$ where there are only transmission constraints at the boundaries of country-zones.

\subsection{Breakdowns by country}

Figures \ref{fig:breakdown} and \ref{fig:breakdown_detail} show the breakdown of total costs by technology for the whole of Europe. However, it could be that for each technology, the spatial distribution is unstable, moving from country to country with the clustering changes.

\begin{figure*} \centering
	\includegraphics[width=\textwidth]{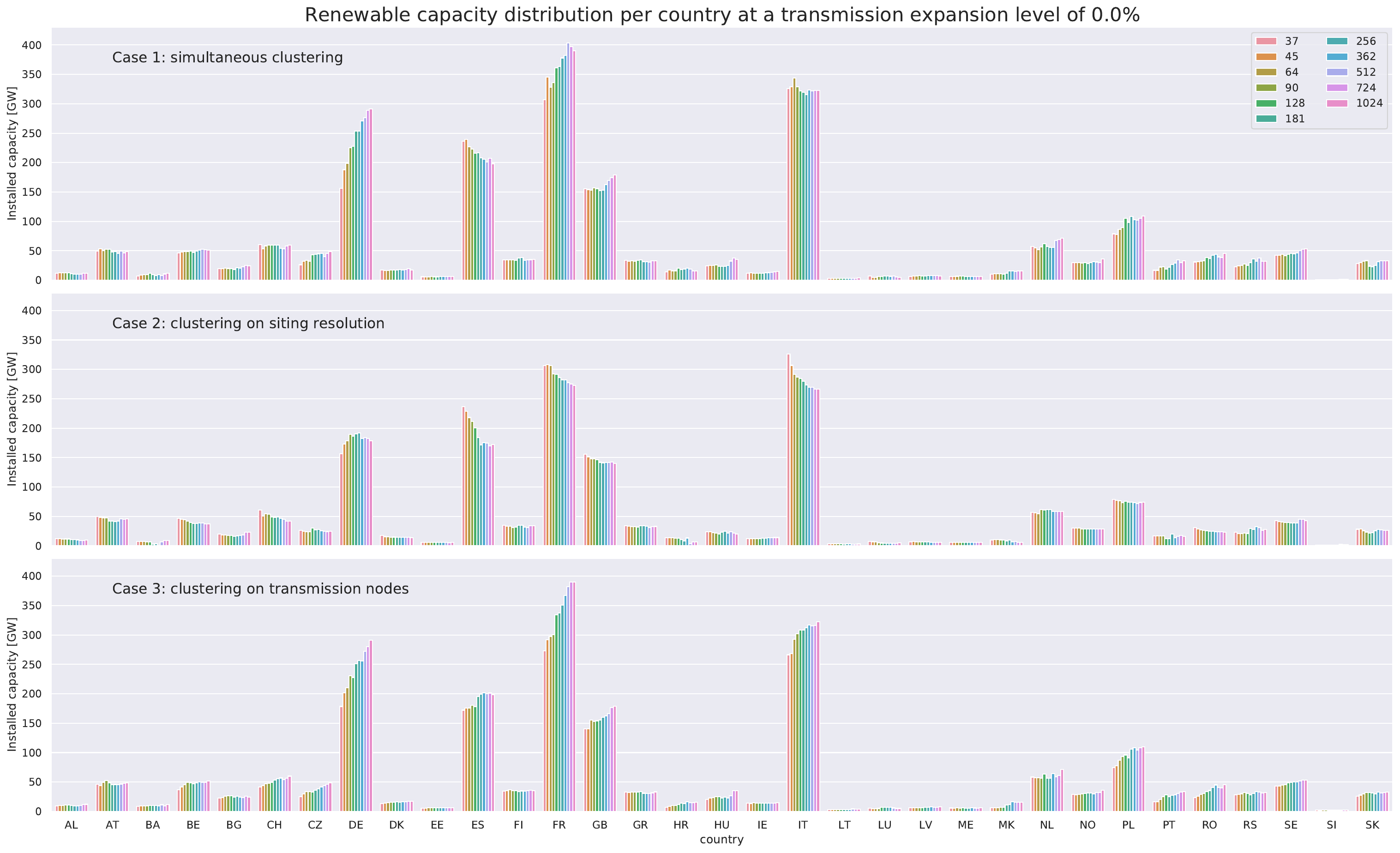}
	\caption{Capacities per country for the no transmission expansion scenario in all three cases.} \label{fig:per_country}
\end{figure*}

For a better understanding of the spatial distribution of installed capacity, we examine the total installed renewable capacity per country in all Cases in Figure \ref{fig:per_country} with no transmission expansion. The general trend is that the total installed capacity per country is relatively stable with cluster resolution. In Case 2 capacity decreases with resolution, since the exploitation of better resource sites means that less capacity is needed for a given energy yield. The opposite effect is seen in Case 3, while Case 1 reveals a mix of the effects of Case 2 and 3.

\subsection{Shadow price of line volume constraint}

The shadow price $\mu_\mathrm{trans}$ of the transmission expansion
constraint in equation \eqref{eq:lvcap} corresponds to the system
cost benefit of an incremental MWkm of line volume. Read another way,
it is the line cost required to obtain the same solution with the
constraint removed (i.e. lifting the constraint into the objective
function as a Lagrangian relaxation).

\begin{figure} \centering
	\includegraphics[width=.45\textwidth]{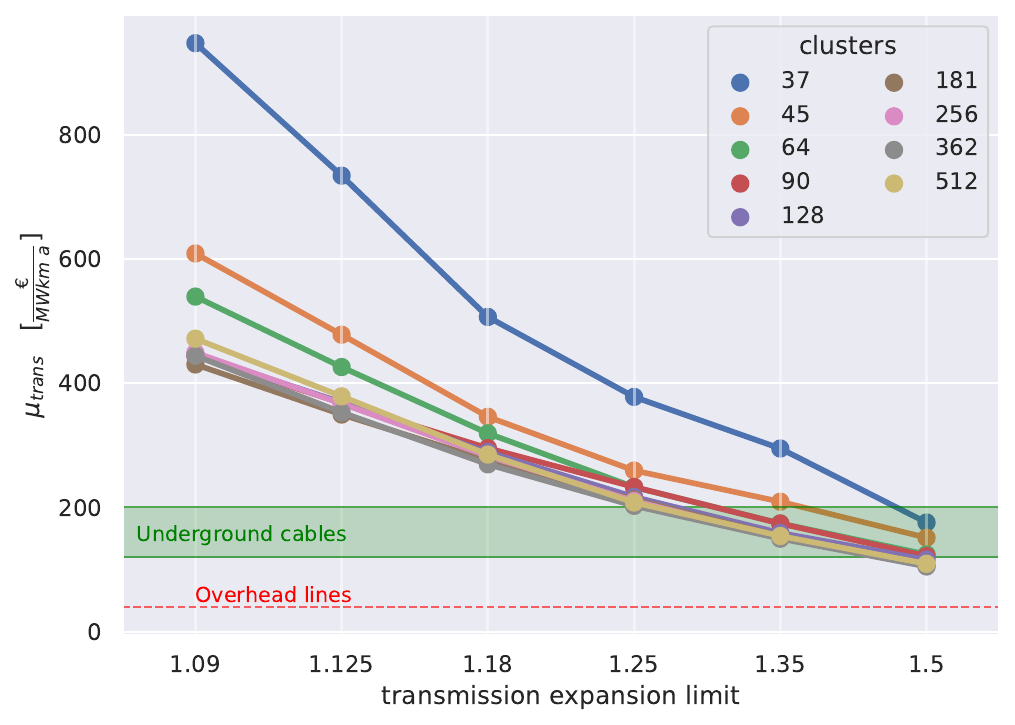}
	\caption{Shadow (dual) price of the line volume constraint.} \label{fig:shadow}
\end{figure}

We present the resulting shadow prices in Figure \ref{fig:shadow}, where they are compared with the annuity for underground and overhead lines.
Using the cost of underground cables, the cost-optimal solution would give a grid expansion of 25-50\% at high resolution. For overhead transmission, the cost optimum would be over 50\%.

\begin{figure*} \centering
	\includegraphics[width=\textwidth]{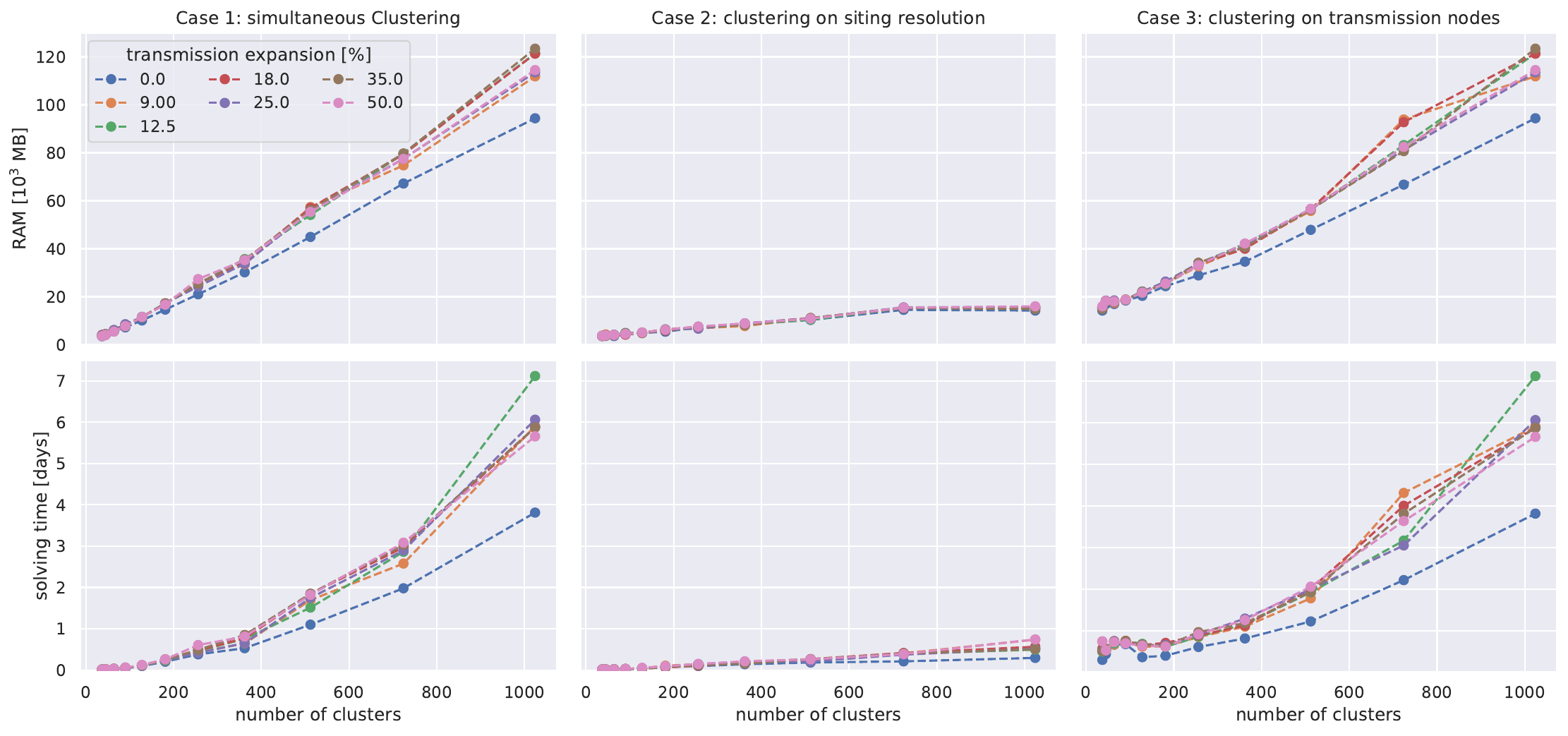}
	\caption{Memory consumption and solving time.} \label{fig:solvingtime}
\end{figure*}

\subsection{Capacity factors within each cluster region for wind and solar}

\begin{figure*} \centering
	\includegraphics[height=.3\textheight]{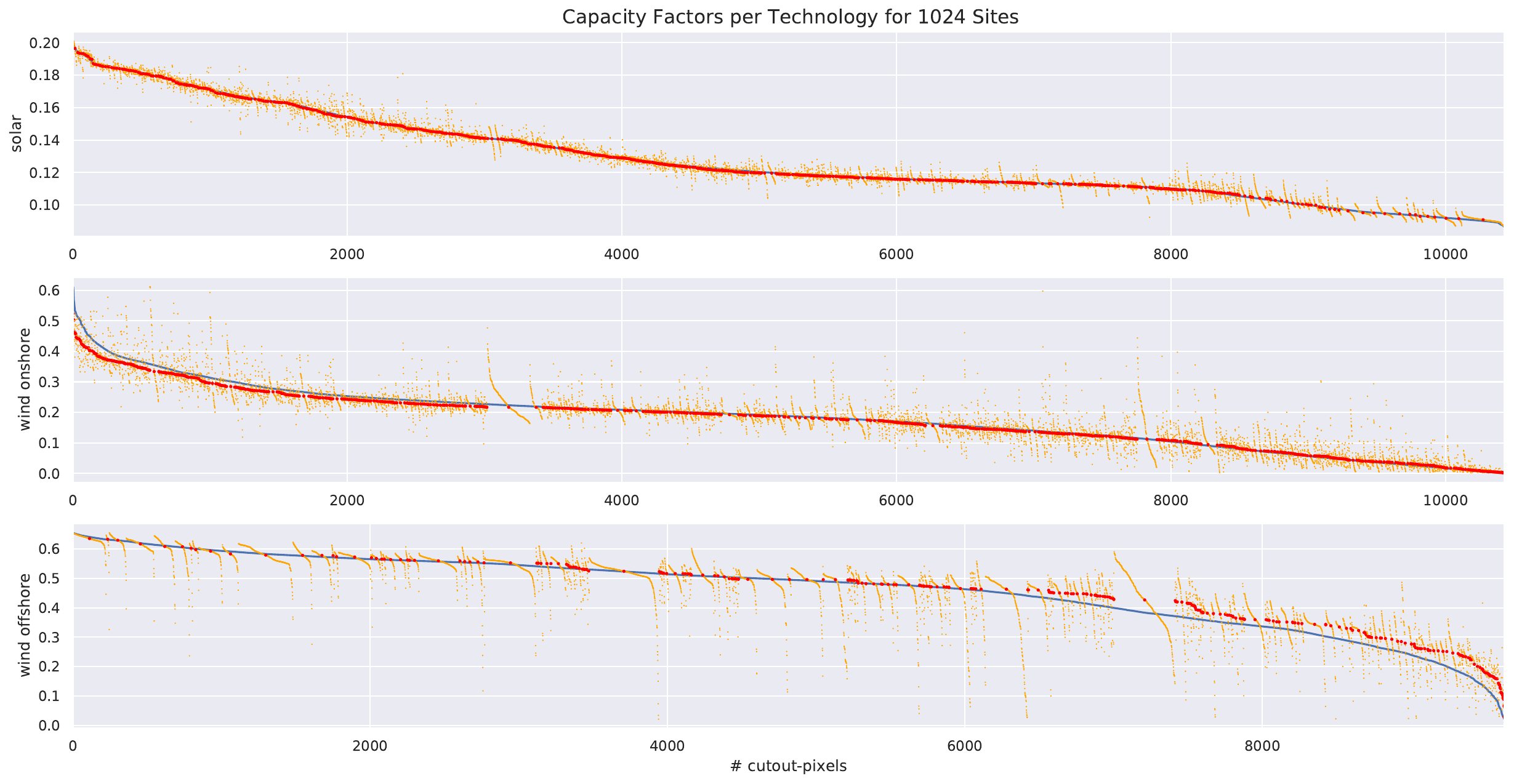}
	\includegraphics[height=.3\textheight]{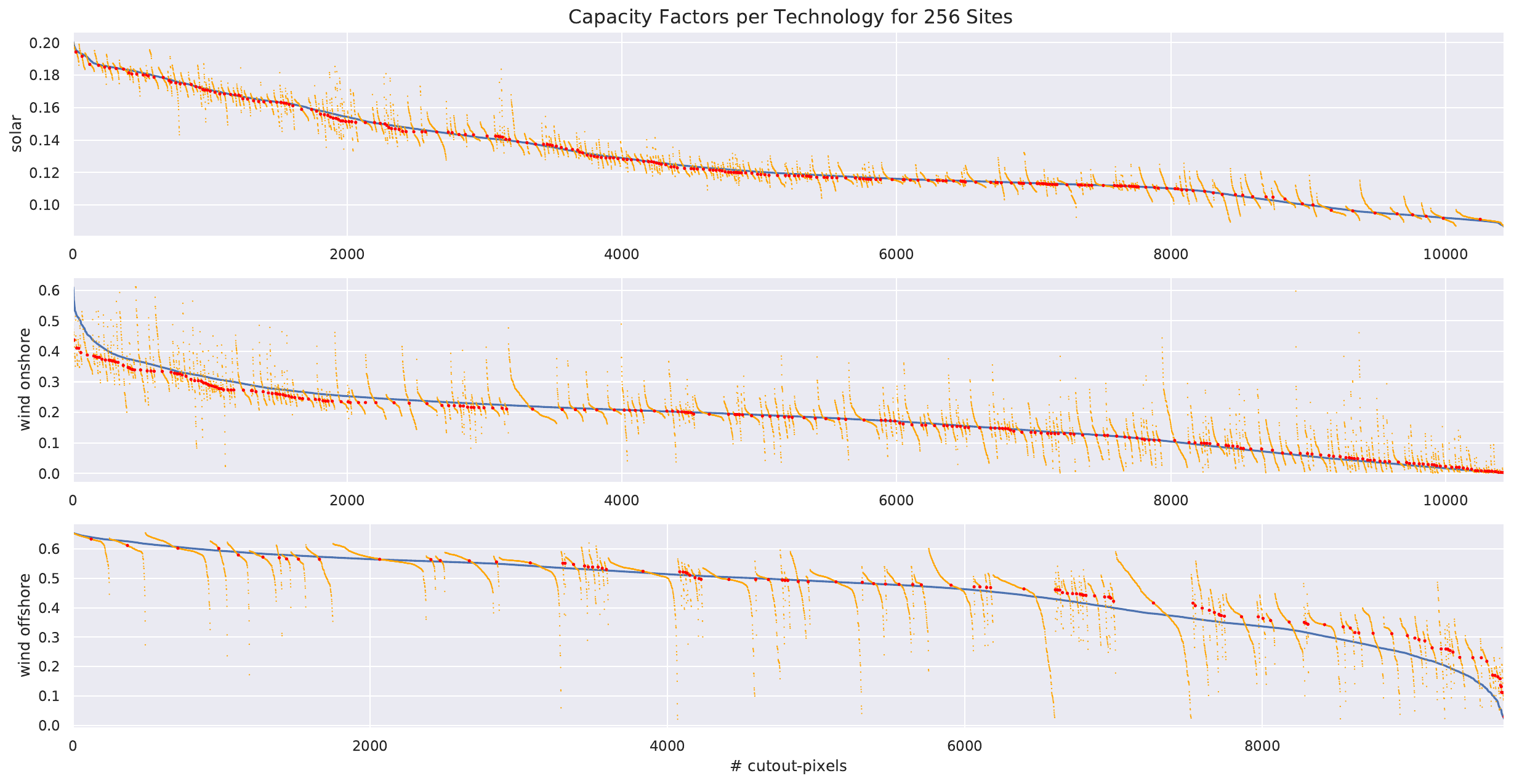}
	\includegraphics[height=.3\textheight]{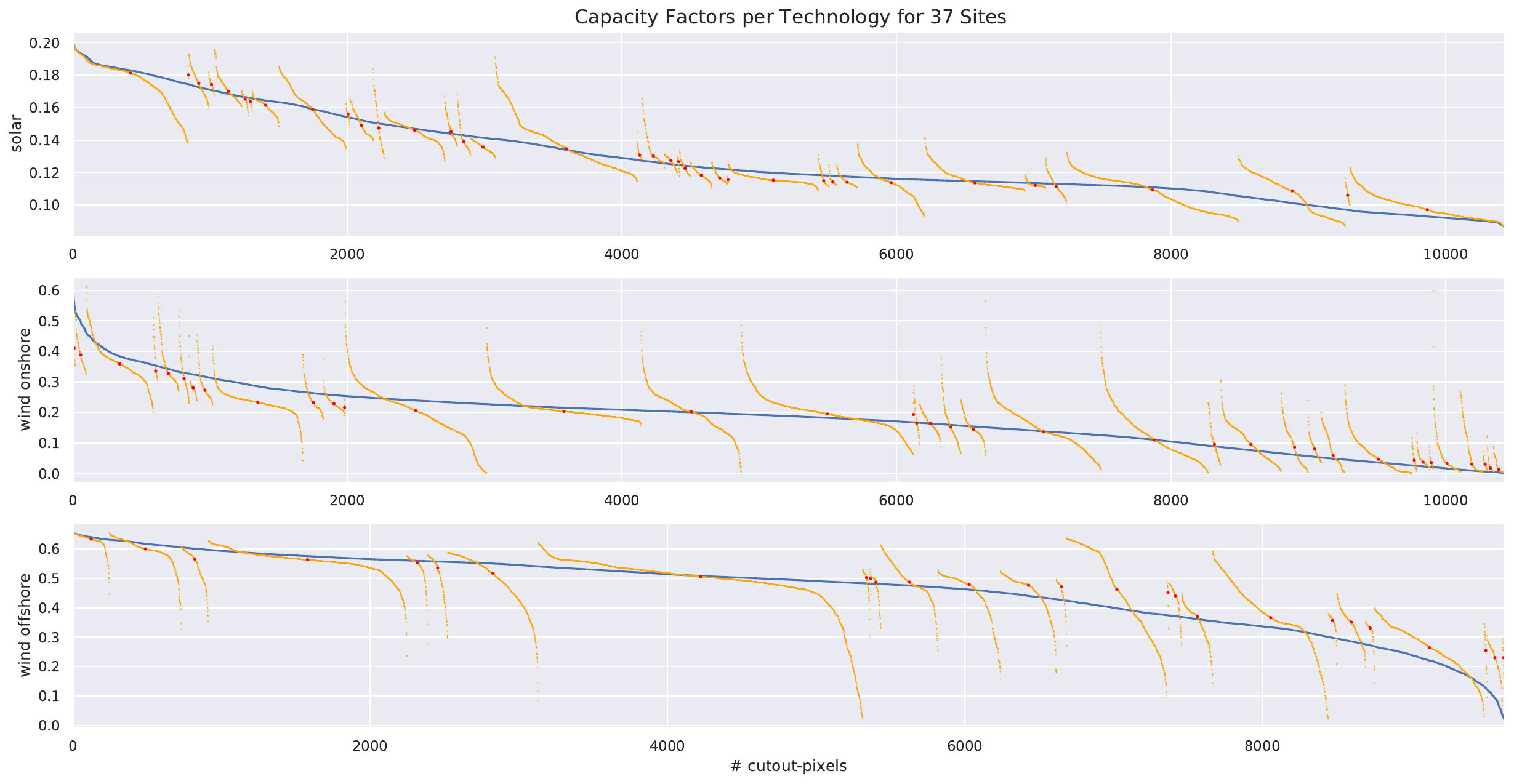}
	\caption{Breakdown of capacity factors per technology for the weather cutout pixels inside each cluster region as a duration curve (orange), with the median marked in red. The overall duration curve of pixel capacity factors for the whole of Europe is plotted in blue.} \label{fig:durationcurves}
\end{figure*}

In this subsection we analyse the homogeneity of time-average capacity
factors for wind and solar within each cluster region as the number of
clusters changes. Duration curves of the capacity factors in each of
the 0.3$^\circ$ $\times$ 0.3$^\circ$ weather pixels of the original
ERA5 reanalysis dataset \cite{ERA5} for the European area (`cutout')
are plotted in blue in Figure \ref{fig:durationcurves}. In addition,
the duration curves for the pixels in each cluster are plotted in
orange, with the median for each cluster in red. This reveals how much
the capacity factors of wind and solar vary within each cluster
region, compared to the whole of Europe.  Table \ref{tab:STDcapfacs} presents the average standard deviation with each cluster region for each technology and resolution.

For a high resolution of 1024 clusters, we observe that the median values
(red dots) for solar lie very close to the representative values of
Europe (black line) with a relatively small average standard deviation of $1.9\cdot 10^{-3}$ inside each cluster region (scattering of the orange dots). In the case of onshore wind, the high capacity factors
are underestimated by the median value, while intermediate and low
capacity factors are represented with a minor difference between
median and representative European value. For onshore wind, the average standard deviation of the
capacity factors within each region is larger than for solar by one magnitude ($\mathcal{O}(10^{-2})$, represented by the scattering of orange dots). The largest variance can be observed in offshore regions, where the average standard deviation is $4.3\cdot 10^{-2}$, twice as large as for onshore regions, and the low capacity factors are overestimated by their representative median values.

In the case of 256 clusters, the standard deviation per region (scattered orange dots) doubles compared to a resolution of $1024$ sites for solar and increases by $\sim 50\%$ for onshore and offshore wind. However, the median values (red dots) per site do not change much compared to the higher resolution case. Only at very low resolutions or, in the extreme, one site representing one country-zone, the median values (red dots) do not agree with the European curve (black line), and the capacity values per site (orange scattered dots) cover a wide range of values (for example $0-0.5$ for wind onshore, or $0.11-0.0.18$ for solar). At 37 nodes, the average standard deviation is three times larger for solar compared to a resolution of 1024 sites and twice as large for onshore wind.

From this analysis we can conclude that a resource resolution of at least several hundred nodes is required to adequately capture the resource variation within Europe, with a higher resolution required for wind than for solar.

\begin{table} \centering
	\begin{tabular}{@{} cccc @{}}
		\toprule
		n clusters & solar & wind onshore & wind offshore \\ \midrule
		$1024$ & $1.9\cdot10^{-3}$ & $2.2\cdot10^{-2}$ & $4.3\cdot10^{-2}$ \\
		$724$ & $2.3\cdot10^{-3}$ & $2.5\cdot10^{-2}$ & $4.5\cdot10^{-2}$\\
		$512$ & $2.7\cdot10^{-3}$ & $2.8\cdot10^{-2}$ & $4.9\cdot10^{-2}$ \\
		$362$ & $3.2\cdot10^{-3}$ & $3.3\cdot10^{-2}$ & $5.1\cdot10^{-2}$ \\
		$256$ & $3.7\cdot10^{-3}$ & $3.6\cdot10^{-2}$ & $5.3\cdot10^{-2}$ \\
		$181$ & $4.2\cdot10^{-3}$ & $3.9\cdot10^{-2}$ & $5.7\cdot10^{-2}$ \\
		$128$ & $4.5\cdot10^{-3}$ & $4.3\cdot10^{-2}$ & $5.8\cdot10^{-2}$ \\
		$90$ & $5.0\cdot10^{-3}$ & $4.6\cdot10^{-2}$ & $5.9\cdot10^{-2}$ \\
		$64$ & $6.1\cdot10^{-3}$ & $4.9\cdot10^{-2}$ & $6.2\cdot10^{-2}$ \\
		$45$ & $6.1\cdot10^{-3}$ & $4.9\cdot10^{-2}$ & $6.2\cdot10^{-2}$ \\
		$37$ & $6.2\cdot10^{-3}$ & $4.9\cdot10^{-2}$ & $6.2\cdot10^{-2}$ \\ \bottomrule
	\end{tabular}
	\caption{average standard deviation of the capacity factor (per unit) per region for a network resolution of $1024$, $256$ and $37$ sites.} \label{tab:STDcapfacs}
\end{table}

\subsection{Limitations of this study}

The need to solve the models at high spatial resolution and 3-hourly
temporal resolution in reasonable time means that compromises have
been made elsewhere: the conventional generation technologies are
limited to hydroelectricity and gas turbines, the storage is limited
to batteries and hydrogen storage, only a single weather year is modelled, and ancillary services, grid losses, discretisation of new grid capacities,
distribution grids and forecast error are not modelled. This allows us
to focus on the main interactions between wind, solar and the
transmission grid; the effects of the other factors are expected to be
small \cite{burdenresponse} since wind and solar investment dominates
system costs. If it were cost-effective to build dispatchable
low-carbon generators like nuclear or fossil generators with carbon
capture and sequestration, then the effects of resource and network
resolution would be dampened, since there would be less wind and solar
investment.

Some of the quantitative conclusions may depend on the technology
assumptions, such as the relative cost of solar PV, onshore wind and
offshore wind. However, investigations of the sensitivities of similar
models to generation costs \cite{Schlachtberger2018} and of the
near-optimal space of solutions \cite{Neumann2019b} have shown that a
large share of wind in low-cost scenarios for Europe is robust across
many scenarios because of the seasonal matching of wind to demand in
Europe. It is the interactions between wind and the transmission grid
that drive the results in this paper.

The results may also change as additional energy sectors are coupled
to the power sector, such as building heating, transport and
non-electric industry demand. While extra flexibility from these
sectors might offer an alternative to grid expansion, grid expansion
is still expected to be cost-effective \cite{Brown2018}, while the
effects of resource resolution on the optimal solution remain the
same.

In the present paper different market structures to today's are assumed,
namely nodal pricing to manage grid congestion, and a high CO$_2$
price to obtain a 95\% CO$_2$ reduction compared to 1990 levels.

We weighted the distribution of wind and solar inside each nodal region (Voronoi
cell) proportional to the installable capacity and capacity factor at
each weather grid cell \cite{PyPSA-Eur}. This means good and bad sites
are not mixed evenly, but skewed slightly towards good sites. This
effect disappears at high resolution, where the capacity factor is
more uniform inside each Voronoi cell.

Another approach would be to keep a low one-node-per-country network
resolution and then have multiple resource classes defined not by
region, like our Case 2, but by capacity factor
\cite{Schlachtberger2017,REICHENBERG2018914,Mattsson2020} (e.g. a good class with sites with full load hours above
2000, a medium class between 1500 and 2000, and a bad class below
1500). This would also be beneficial but would not be compatible with
the increasing grid resolution, since the generators in each class
would be spread non-contiguously over the country.

%% Here is the endmatter stuff: Supplementary Info, etc.
%% Use \item's to separate, default label is "Acknowledgements"

%\begin{addendum}

%\end{addendum}

%\begin{align*}
%x_c &= \frac{1}{|N_c|}\sum_{i\in N_c} x_i \,, \\
%v_{\mathrm{nom},\,c} &= \mathrm{max}_{i\in N_c} v_{\mathrm{nom},\,i}\,, \\
%v_{\mathrm{mag}_\mathrm{pu,\, max},c} &= \mathrm{min}_{i\in %N_c}v_{\mathrm{mag}_\mathrm{pu,\, max},i}\,, \\
%v_{\mathrm{mag}_\mathrm{pu,\, min},\,c} &= \mathrm{max}_{i\in %N_c}v_{\mathrm{mag}_\mathrm{pu,\, min},i}\,,
%\end{align*}

%\begin{align*}
%c_{i,s} &\mapsto w_i \cdot c_{i,s} \\
%p_{\mathrm{nom}_\mathrm{max},i} &\mapsto \frac{1}{w_i} min \\
%w_i &\mapsto \sum \\
%p_{\mathrm{nom},i} &\mapsto \sum \\
%c_{i,s} &\mapsto \sum
%\end{align*}

\begin{table*} \centering
	\caption{Aggregation rules for attributes of nodes and attached assets} \label{tab:nodes}
	\begin{tabular}{@{} llll @{}}
		\toprule
		attribute & aggregated attribute & mapping & values or units \\
		\midrule
		latitude \& longitude  & $x_c$ & $\frac{1}{|N_c|}\sum_{i\in N_c} x_i$ & $\mathbb{R}^2$ \\
		%$v_{\mathrm{nom}_c}$ & $\mathrm{max}_{i\in N_c} v_{\mathrm{nom}_i}$ & $kV$, nominal voltage\\
		%$v_{\mathrm{mag}_\mathrm{max}c}$ & $\mathrm{min}_{i\in N_c}v_{\mathrm{mag}_\mathrm{max}i}$ & per unit \\
		%$v_{\mathrm{mag}_\mathrm{min}c}$ & $\mathrm{max}_{i\in N_c}v_{\mathrm{mag}_\mathrm{min}i}$ & per unit \\
		\hline
		%asset capital cost & $c_{c,s}$ & $\sum_{i\in N_c} \frac{w_{i,s}}{w_{c,s}}\cdot c_{i,s}$ & $\frac{\euro}{MW}$ \\
		(optimal) power capacity & $G_{c,s}$ & $\sum_{i\in N_c} G_{i,s}$ & $MW$ \\
		asset installable potential & $G^\mathrm{max}_{c,s}$ & $\sum_{i\in N_c} G^\mathrm{max}_{i,s}$ & $MW$ \\
		\bottomrule
	\end{tabular}
\end{table*}

%{'weight': np.sum}$

%\begin{align*}
%l^{c,d} &\mapsto \sum_{(i,j)\in N_{c,d}} l^{i,j} \\
%p_\mathrm{nom}^{c,d} &\mapsto \min_{(i,j)\in N_{c,d}}\{p_\mathrm{nom}^{i,j}\} \\
%\mathrm{frac}_{\mathrm{underwater}}^{c,d} &\mapsto \frac{1}{l^{c,d}}\sum_{(i,j)\in %N_{c,d}}l^{i,j}\cdot\mathrm{frac}_{\mathrm{underwater}}^{i,j}\,,
%\end{align*}
%where $l$ denotes the length of a link, $p_{\mathrm{nom}}$ is the limit of active power which can pass through and frac$_\mathrm{underwater}$ denotes the length under water in percent. Here, superscript indices refer to the nodes connected by a link. For simplicity we will omit them in the following whenever they are not needed and it is clear which line is meant. Aggregated links remain undirected with a maximal and minimal per unit dispatch of $\pm 1$.

\begin{table*}\centering
	\caption{Aggregation rules for attributes of lines in series} \label{tab:linesseries}
	\begin{tabular}{@{} llll @{}}
		\toprule
		attribute & aggregated attribute & mapping & values or units \\
		\midrule
		length (HVDC lines) & $l_{c,d}$ & $\min_{\ell_{i,j}\in N_{c,d}} l_{i,j}$ & km \\
		power capacity & $F_{c,d}$ & $\sum_{\ell_{i,j}\in N_{c,d}}F_{i,j}$ & MVA\\
		fraction of length underwater & $u_{c,d}$ & $\frac{1}{l_{c,d}}\sum_{\ell_{i,j}\in N_{c,d}}l_{i,j}\cdot u_{i,j}$ & per unit \\
		%$p_\mathrm{min\backslash max}$ & $\pm1$ & per unit \\
		\bottomrule
	\end{tabular}
\end{table*}

%\begin{align*}
%R &= \frac{1}{\sum_{l\in N_c} \frac{v_f}{l_f \cdot R_l}} \\
%x &= \frac{1}{\sum_{l \in N_c} \frac{v_f}{l_f \cdot x_l}} \\
%g &= \sum_{l\in N_c} v_f \cdot l_f \cdot g_l \\
%b &= \sum_{l\in N_c} v_f \cdot l_f \cdot b_l \\
%\mathrm{terr}_f &= \frac{1}{|N_c|}\sum_{l\in N_c} \mathrm{terr}_{f,l} \\
%s_\mathrm{nom} &= \sum_{l\in N_c} s_{\mathrm{nom},l} \\
%s_{\mathrm{nom}_\mathrm{min}} &= \sum_{l\in N_c} s_{\mathrm{nom}_\mathrm{min},l} \\
%s_{\mathrm{nom}_\mathrm{max}} &= \sum_{l\in N_c} s_{\mathrm{nom}_\mathrm{max},l}\\
%n_\mathrm{parallel} &= \sum_{n\in N_c} n_{\mathrm{parallel},l} \\
%c_l &= l_f \cdot \sum_{l\in N_c} \|s_{\mathrm{nom},l}\|_1 \cdot c_l \\
%v_{\mathrm{ang}_\mathrm{min}} &= \mathrm{max}_{l\in N_c} %v_{\mathrm{ang}_\mathrm{min}} \\
%v_{\mathrm{ang}_\mathrm{max}} &= \mathrm{min}_{l\in N_c} %v_{\mathrm{ang}_\mathrm{max}}
%\end{align*}

\begin{table*}\centering
	\caption{Aggregation rules for attributes of lines in parallel}         \label{tab:linesparallel}
	\begin{tabular}{@{} llll @{}}
		\toprule
		attribute & aggregated attribute & mapping & values or units \\
		\midrule
		%line length ratio &  $r_{c,d}$ & $\frac{r_{c,d}}{r_\ell} \quad \forall \ell\in N_{c,d}$ & $\mathbb{R}^{|N_{c,d}|}$ \\
		%$R_{c,d}$ & $\left(\sum_{l\in N_{c,d}} \frac{1}{\mathrm{fac}_l \cdot R_l}\right)^{-1}$ & $\Omega$, resistance \\
		%$x_{c,d}$ & $\left(\sum_{l \in N_{c,d}} \frac{1}{\mathrm{fac}_l \cdot x_l}\right)^{-1}$ & $\Omega$, reactance \\
		%$g_{c,d}$ & $\sum_{l\in N_{c,d}} \mathrm{fac}_l \cdot g_l$ & $S$, conductance\\
		%$b$ & $\sum_{l\in N_{c,d}} \mathrm{fac}_l \cdot b_l$ & $S$, susceptance \\
		power capacity & $s^\mathrm{nom}_{c,d}$ & $\sum_{\ell_{i,j}\in N_{c,d}} s^\mathrm{nom}_{i,j}$ & $MVA$ \\
		power capacity maximum & $s^\mathrm{min}_{c,d}$ & $\sum_{\ell_{i,j}\in N_{c,d}} s^\mathrm{min}_{i,j}$ & $MVA$\\
		power capacity minimum & $s^\mathrm{max}_{c,d}$ & $\sum_{\ell_{i,j}\in N_{c,d}} s^\mathrm{max}_{i,j}$ & $MVA$ \\
		number of parallel lines & $n^\mathrm{parallel}_{c,d}$ & $\sum_{\ell_{i,j}\in N_{c,d}} n^\mathrm{parallel}_{i,j}$ & $\mathbb{R}$ \\
		terrain factor for capital costs & $\mathrm{terr}_{c,d}$ & $\frac{1}{|N_{c,d}|}\sum_{\ell_{i,j}\in N_{c,d}} \mathrm{terr}_{i,j}$ & per unit \\
		%capital cost & $c_{c,d}$ & $ \sum_{\ell_{i,j}\in N_{c,d}} r_{i,j} \cdot \|s^\mathrm{nom}_{i,j}\|_1 \cdot c_{i,j}$ & $\frac{\euro}{MVA}$ \\
		%$v_{\mathrm{ang}_\mathrm{min}c,d}$ & $\mathrm{max}_{l\in N_{c,d}}$ $v_{\mathrm{ang}_\mathrm{min}l}$ & $^\circ$, degrees \\
		%$v_{\mathrm{ang}_\mathrm{max}c,d}$ & $\mathrm{min}_{l\in N_{c,d}} v_{\mathrm{ang}_\mathrm{max}l}$ & $^\circ$, degrees
		\bottomrule
	\end{tabular}
\end{table*}

\printcredits

\bibliographystyle{cas-model2-names}

\bibliography{spatial}


\end{document}